\documentclass[a4paper,11pt]{article}
\usepackage[utf8]{inputenc}
\usepackage[english]{babel}

\usepackage{authblk}
\usepackage{setspace}
\usepackage[margin=1.25in]{geometry}
\usepackage{graphicx}
\graphicspath{ {./figures/} }
\usepackage{subcaption}
\usepackage{amsmath}
\usepackage[switch]{lineno} 
\usepackage{amssymb}
\usepackage{xspace}
\usepackage{xcolor}
\usepackage{csquotes}
\usepackage{comment}
\usepackage{multirow}

\newcommand{\MADGRAPHfive}{\textsc{MadGraph5}\xspace}
\newcommand{\GEANTfour}{\textsc{Geant4}\xspace}
\newcommand{\PTARMIGAN}{\textsc{Ptarmigan}\xspace}
\newcommand{\xzero}{\,X_0\xspace}

\newcommand{\MeV}{~\mathrm{MeV}\xspace}
\newcommand{\GeV}{~\mathrm{GeV}\xspace}

\newcommand{\TW}{~\mathrm{TW}\xspace}

\newcommand{\ns}{~\mathrm{ns}\xspace}

\newcommand{\mum}{~\mathrm{\mu m}\xspace}
\newcommand{\mm}{~\mathrm{mm}\xspace}
\newcommand{\cm}{~\mathrm{cm}\xspace}
\newcommand{\m}{~\mathrm{m}\xspace}

\newcommand{\T}{~\mathrm{T}\xspace}
\newcommand{\mx}{~m_{X}\xspace}

\let\originalLesssim\lesssim
\renewcommand{\lesssim}{~\originalLesssim\xspace}

\let\originalTextgreater\textgreater
\renewcommand{\textgreater}{~\originalTextgreater\xspace}

\usepackage{soul}

\let\oldsim\sim 
\renewcommand{\sim}{{\oldsim}}
\usepackage[style=ieee, 
citestyle=numeric-comp,
sorting=none]{biblatex}
\title{Layout optimization for the LUXE-NPOD experiment}

\author[3]{Melissa Almanza Soto}
\author[2]{Oleksandr Borysov}
\author[1]{Torben Ferber}
\author[3]{Shan Huang}
\author[3]{Adrián Irles}
\author[1]{Markus Klute}
\author[3]{Jes\'us P. M\'arquez Hern\'andez}
\author[3]{Josep P\'erez Segura}
\author[1]{Raquel Quishpe}
\author[4]{Yotam Soreq}
\author[2]{Noam Tal Hod}
\author[1]{Nicol\`o Trevisani}

\affil[1]{Institute for Experimental Particle Physics (ETP), Karlsruhe Institute of Technology (KIT), D-76131 Karlsruhe, Germany}
\affil[2]{Department of Particle Physics and Astrophysics,
Weizmann Institute of Science, Rehovot 7610001, Israel}
\affil[3]{Instituto de F\'isica Corpuscular (IFIC), CSIC-Universitat de Val\`encia, 46980 Paterna, Spain}
\affil[4]{Physics Department, Technion - Israel Institute of Technology, Haifa 3200003, Israel}

\date{}

\begin{document}



\maketitle
\begin{abstract} 
    Beam dump experiments represent an effective way to probe new physics in a parameter space, where new particles have feeble couplings to the Standard Model sector and masses below the GeV scale. 
    The LUXE experiment, designed primarily to study strong-field quantum electrodynamics, can be used also as a photon beam dump experiment with a unique reach for new spin-0 particles in the $10-350$~MeV mass and $10^{-6}-10^{-3}~\GeV^{-1}$ couplings to photons ranges.
    This is achieved via the ``New Physics search with Optical Dump'' (NPOD) concept. 
    While prior estimations were obtained with a simplified model of the experimental setup, in this work we present a systematic study of the new physics reach in the full, realistic experimental apparatus, including an existing detector to be used in the LUXE NPOD context.
    We furthermore investigate updated scenarios of LUXE's experimental plan and confirm that our results are in agreement with the original estimations of a background-free operation.

\end{abstract}


\section{Introduction}
\label{sec:intro}

The Standard Model~(SM) of particle physics has passed decades-long rigorous testing through numerous experiments.
While it has successfully predicted experimental results across a wide range of energy scales, it is widely acknowledged that the SM cannot describe all the observed phenomena in nature.
Observations such as neutrino oscillations, cosmological baryon asymmetry, and the substantial body of evidence supporting the existence of dark matter lie beyond the SM explanatory power.
Despite continuous efforts to uncover physics beyond the SM~(BSM), unambiguous evidence has remained elusive, prompting the need for innovative experimental approaches to probe New Physics~(NP).

Beam dump experiments are ideal testbeds for directly exploring various NP models 
in a parameter space not covered by colliders~\cite{PhysRevD.38.3375, BF01548556}.
A multitude of such experiments is currently taking data or is in the design phase, each targeting distinct signatures or regions of the parameter space.
A common characteristic that these experiments share is the usage of an intense beam of SM particles (typically protons, electrons, or photons), a thick, solid, high-$Z$ material beam dump, a decay volume, and a detector.
In these setups, the incoming particles 
interact with the beam dump material, resulting in the production of light NP particles that may be highly boosted, feebly interacting, and long-lived.
The feebly-interacting particles~(FIPs)~\cite{Beacham_2019} freely propagate through the dump material. 
If they decay to SM particles within the decay volume, their decay products may interact with the detector.
The beam dump is primarily designed to absorb primary beam particles and the secondary particles that are produced in the electromagnetic and hadronic showers generated by it.

The LUXE experiment~\cite{abramowicz2023technical} at the Eu.XFEL~\cite{Altarelli:2006zza} (DESY, Hamburg), is designed to explore strong-field quantum electrodynamics~(SF-QED) by colliding high-energy electrons from the Eu.XFEL accelerator with intense laser pulses.
Due to the high photon flux, the design of the experiment makes it additionally sensitive to the production of new spin-0 particles via photon-dump interaction, as suggested in the first New Physics at Optical Dump~(NPOD) study~\cite{Bai_2022}.
Hereafter, we focus on the pseudo-scalar axion-like particles~(ALPs) case, which is experimentally identical to the scalar case. See \cite{PhysRevLett.123.051103, annurev-nucl-102014-022120, hook2023tasilecturesstrongcp, Irastorza_2018, annurev-nucl-120720-031147} for comprehensive reviews.
As discussed in~\cite{abramowicz2023technical,Bai_2022}, the electron-laser ($e$-laser) interaction generates an intense photon beam via non-linear Compton scattering~(NCS), with a distinct energy spectrum~\cite{Bai_2022}, where strikingly, there could be several ($\sim2-3$ on average) photons produced with GeV-scale energy generated per primary beam electron. 
These NCS photons are tightly collimated along the trajectory of the incoming electrons. 
Therefore, if all primary beam electrons (those that have interacted or not) are effectively dumped after the $e$-laser interaction point, one is left with a pure, ultra-intense GeV-scale photon beam unparalleled to any existing photon facility worldwide.
In the NPOD design, the NCS photons traverse the experimental setup (in vacuum) and reach a physical dump at its end, where their interaction with the material nuclei leads to the production of ALPs.
The decay volume following the beam dump ends with a detector designed to allow the reconstruction of the two photons resulting from the decay of the ALPs.

In this work, we present the outcome of an optimization study, aimed at (i)~validating the original NPOD estimations in the realistic experimental setup, (ii)~potentially enhancing the sensitivity reach of the LUXE BSM program, and (iii)~studying updated scenarios in view of LUXE's updated experimental plan. 
The latter scenarios are based on the recent developments discussed in both the LUXE technical design report~\cite{abramowicz2023technical} and the related ``ELBEX'' beam-extraction EU-funded project~\cite{ELBEX}.
The experimental model is now fully simulated, including all the elements of the LUXE apparatus and the experimental cavern.
The dump design is carefully optimized in all its aspects (depth, radius, and material) to enhance the signal sensitivity, while attempting to maintain a background-free environment.
The possibility of immersing the dump core in a strong magnetic field to 
de-focus the electromagnetic and hadronic showers around the beam axis and hence reduce the collinear, signal-like background components is also explored.
Compared to the original NPOD study~\cite{Bai_2022}, in this work, we include a full detector simulation, which considers the realistic model of the LUXE electromagnetic calorimeter, LUXE ECAL-E, developed by the CALICE Collaboration~\cite{Kawagoe:2019dzh, Breton:2020xel}. 
Furthermore, we include a proper reconstruction and identification analysis.
This detector is already at hand and it can be installed together with LUXE.

This paper is organized as follows: in Sec.~\ref{sec:npodsensitivity} we study the impact of different dump length, decay volume, and detector feature assumptions on the sensitivity; in Sec.~\ref{sec:background} we perform an optimization study aiming to reach a zero-background environment assuming the scenario with the most intense NCS photon beam achievable in LUXE's lifetime; Sec.~\ref{sec:detector} presents the requirements for the detector and the off-the-shelf hardware already available for LUXE. In Sec.~\ref{sec:results} we
present the results of the study. Finally, in Sec.~\ref{sec:outlook} we conclude the discussion and look at future perspectives.

\section{Sensitivity optimization}
\label{sec:npodsensitivity}

\begin{figure}[htbp]
    \centering
    \includegraphics[width=0.8\textwidth]{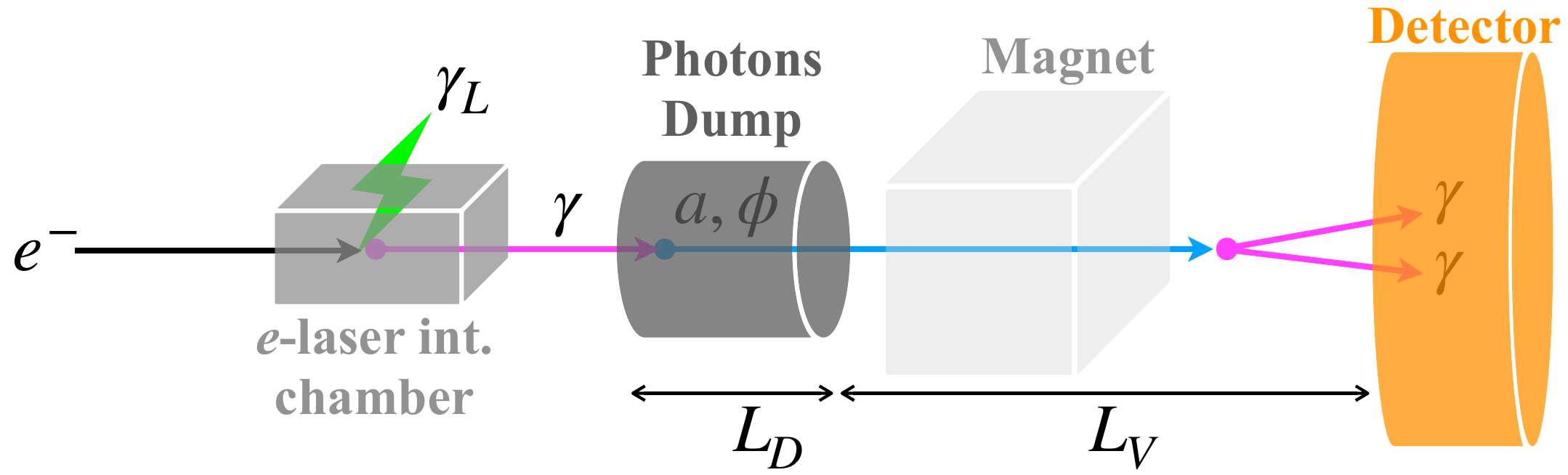}
    \caption{Schematic representation of the LUXE-NPOD concept, showing the NP production. Figure adapted from~\cite{Bai_2022}.
    }
    \label{fig:npod_cartoon}
\end{figure}

LUXE-NPOD focuses on processes involving models, where effective pseudo-scalar and scalar particles (ALPs), denoted as $X=a,\phi$, respectively, can couple to two photons.
The effective interactions can be written as
\begin{align}
    \mathcal{L}_{a,\phi}
    =
    \frac{a}{4\Lambda_a}\tilde{F}_{\mu\nu}F^{\mu\nu}+
    \frac{\phi}{4\Lambda_\phi}F_{\mu\nu}F^{\mu\nu}\, , 
\end{align}
where $F_{\mu\nu}$ is the photon field strength and $\tilde{F}_{\mu\nu}=(1/2)\epsilon_{\mu\nu\alpha\beta}F^{\alpha\beta}$ is its dual.
The ALPs are produced by the interaction of NCS photons with the dump material, via the Primakoff mechanism~\cite{PhysRevLett.123.071801, PhysRevD.34.1326}, and then decay at some point after they exit the dump to two photons.
Fig.~\ref{fig:npod_cartoon} shows a schematic representation of the NP production at LUXE-NPOD. 

The expected number of $X$ produced and detected in the LUXE-NPOD setup is approximated following the conventions and assumptions of~\cite{Bai_2022}. 
At the Eu.XFEL, each bunch contains $N_{e}=1.5\times 10^{9}$ electrons with an energy of $E_{e}=16.5\GeV$~\cite{abramowicz2023technical}. 
We assume one year of data taking, which corresponds to $10^7$ effective live seconds of the experiment.
Assuming an electron-laser collision rate of 1~Hz\footnote{Limited by the repetition rate of the laser.}, we are left with a total number of collisions (bunch crossings, BX) in one year of $N_{\rm BX} = 10^7$.
In the original NPOD study~\cite{Bai_2022} it was found that tungsten is an excellent material to enhance the NP production, while the dump can be kept short as it has a smaller radiation length ($X_0$) compared to other materials like lead or iron.
Therefore, a tungsten dump with a length $L_{D} = 1.0\m$ ($286 \xzero$) is considered as the benchmark in the rest of this work. 
The NCS photon flux is simulated using \PTARMIGAN~v1.4~\cite{Blackburn_2023,Blackburn_2021} for estimating the energy distribution $\mathrm{d} N_{\gamma}/\mathrm{d} E_{\gamma}$.

LUXE will initially run a laser with a power of $10 \TW$ during its phase-0 (phase-0.10).
This will be achieved with the JETI40~\cite{JETI40} laser that has been moved from JENA to DESY in 2025 and is now being recommissioned.
Furthermore, the extraction line and dump for this setup are already designed and budgeted, mostly through the ELBEX project~\cite{ELBEX}.
The detector hardware is also existing at hand as discussed in Sec.~\ref{sec:detector}.
In a later stage, it is foreseen that LUXE phase-0 can run up to a power of $40\TW$ (phase-0.40) with JETI40, whereas in phase-1 a new laser system is planned for a $350\TW$ benchmark.
As shown in~\cite{Bai_2022}, for LUXE phase-0, the photon flux is much smaller than in phase-1, leading to smaller signal yields, but also much smaller background yields, allowing for a shorter dump in phase-0.

The largest flux of NCS photons for the phase-0.10, phase-0.40 and phase-1 setups is obtained using a laser pulse length of 30, 25 and 120~fs and transverse spot size of 5, 6.5 and 10$\mum$, respectively.
To ensure a background-free environment, we therefore consider the laser parameters expected for LUXE phase-1~\cite{abramowicz2023technical} for the dump optimization study.

The ALPs production by the NCS photons in the dump is simulated using \MADGRAPHfive v2.8.1~\cite{Alwall_2011}, with a UFO model~\cite{Degrande_2012} to describe the Primakoff production, and taking into account form factors~\cite{Chen_2017,RevModPhys.46.815,Bjorken_2009}. 
Signal events are simulated for various ALP masses $0.01 ~<~ m_a ~<~ 1.0 ~\GeV/c^2$, and various lifetimes $3.16 \times 10^2 ~< ~c \tau ~< ~3.16 \times 10^6 ~\cm$ in variable steps.
In the simulation, the NP particles decay instantaneously into a pair of photons.
Hence, to simulate the distance, which the $X$ travels before decaying, 
a random length parameter $L_X ~\equiv ~ c\tau_X p_X / m_X$ is drawn from the exponential distribution defined by the particular lifetime $\tau_X$ and momentum $p_X ~ \approx ~ \sqrt{E_{\gamma}^2-m_X^2}$.

\begin{figure}[!t]
     \centering
     \begin{subfigure}[b]{0.32\textwidth}
        \centering
        \includegraphics[width=\textwidth]{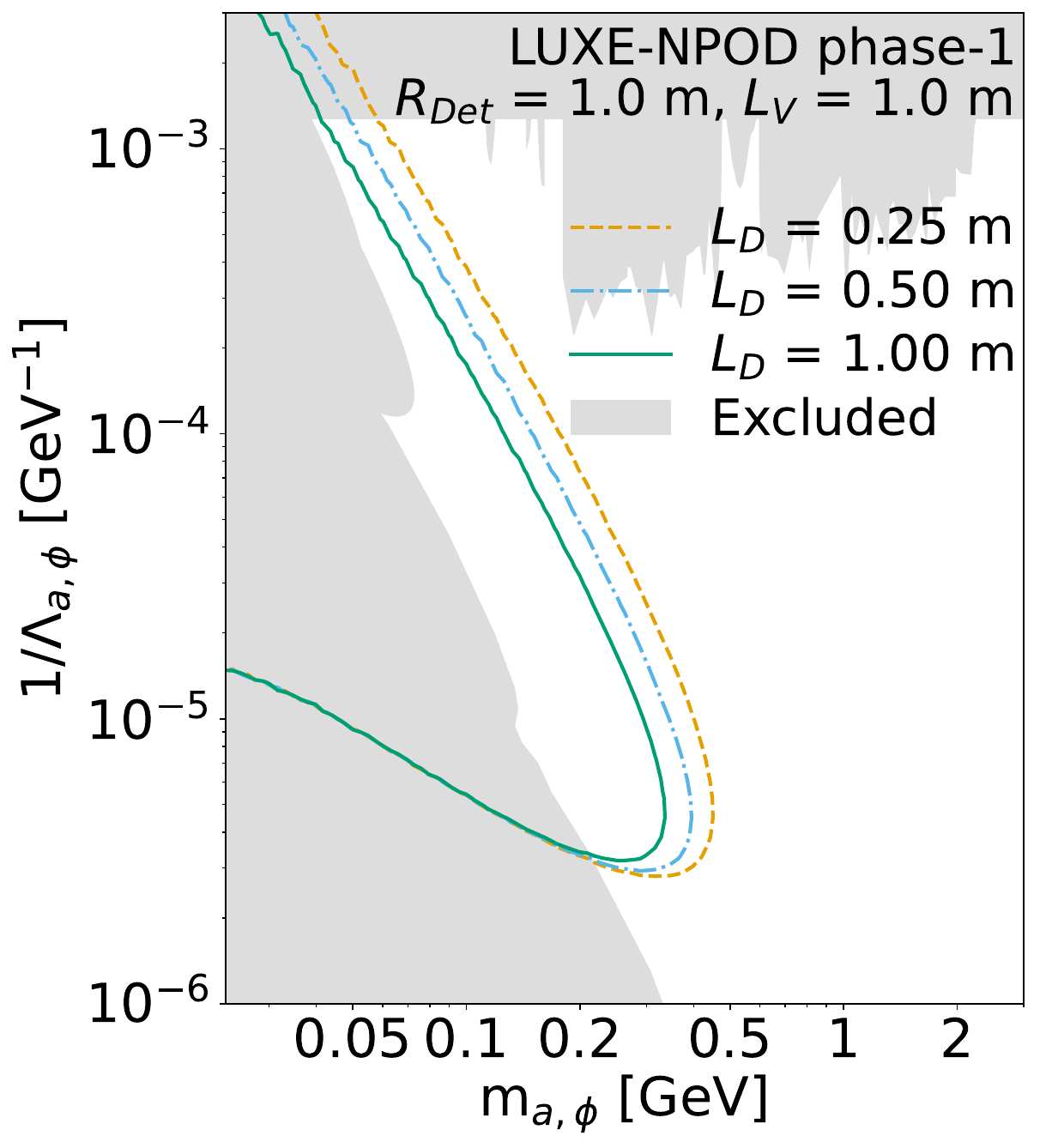}
        \caption{}
        \label{subfig:phase1_sensitivity_dumplength}
    \end{subfigure}
    \hfill
    \begin{subfigure}[b]{0.32\textwidth}
        \centering
        \includegraphics[width=\textwidth]{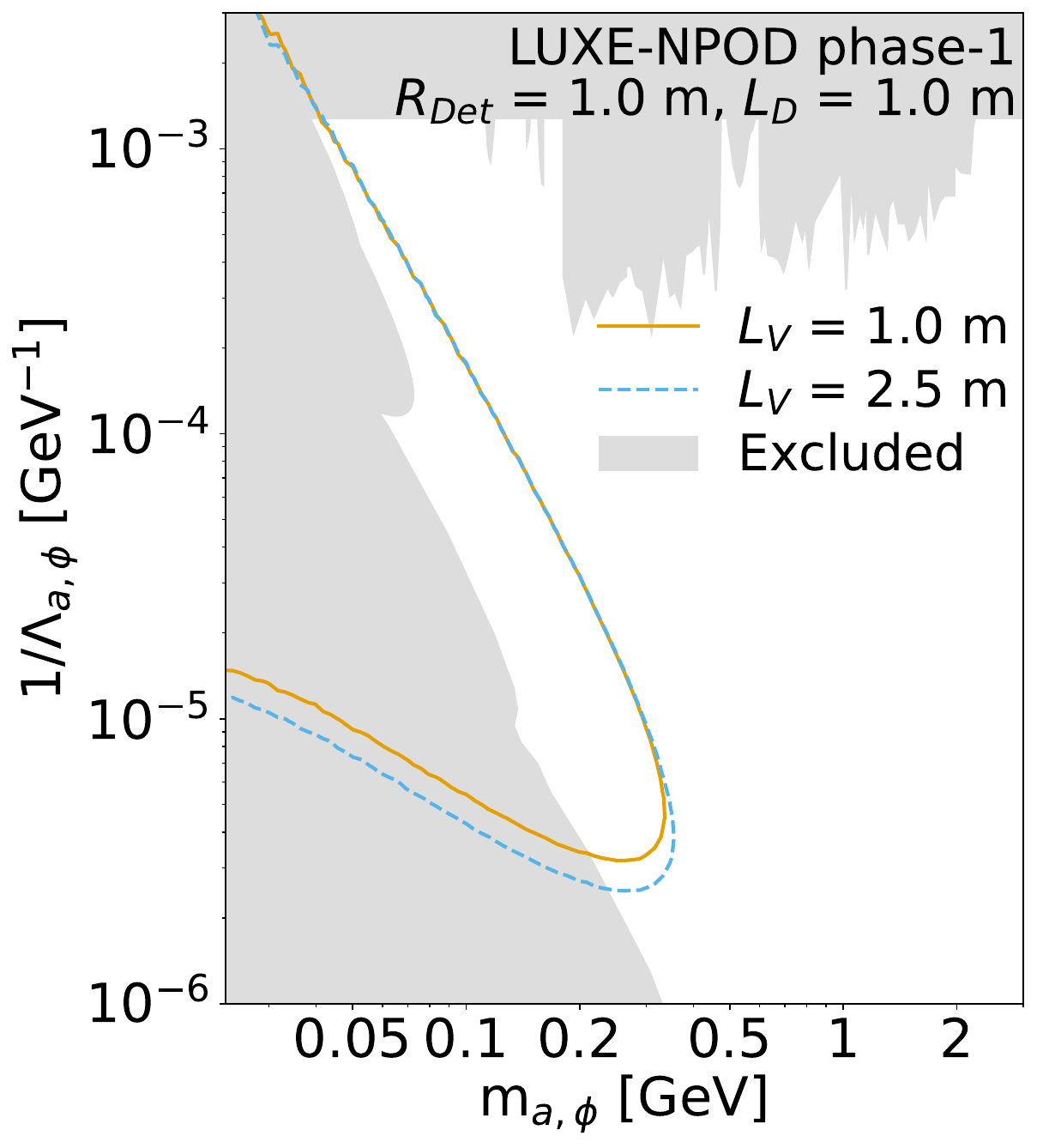}
        \caption{}
        \label{subfig:phase1_sensitivity_decayvolume}
    \end{subfigure}
    \hfill
    \begin{subfigure}[b]{0.32\textwidth}
        \centering
        \includegraphics[width=\textwidth]{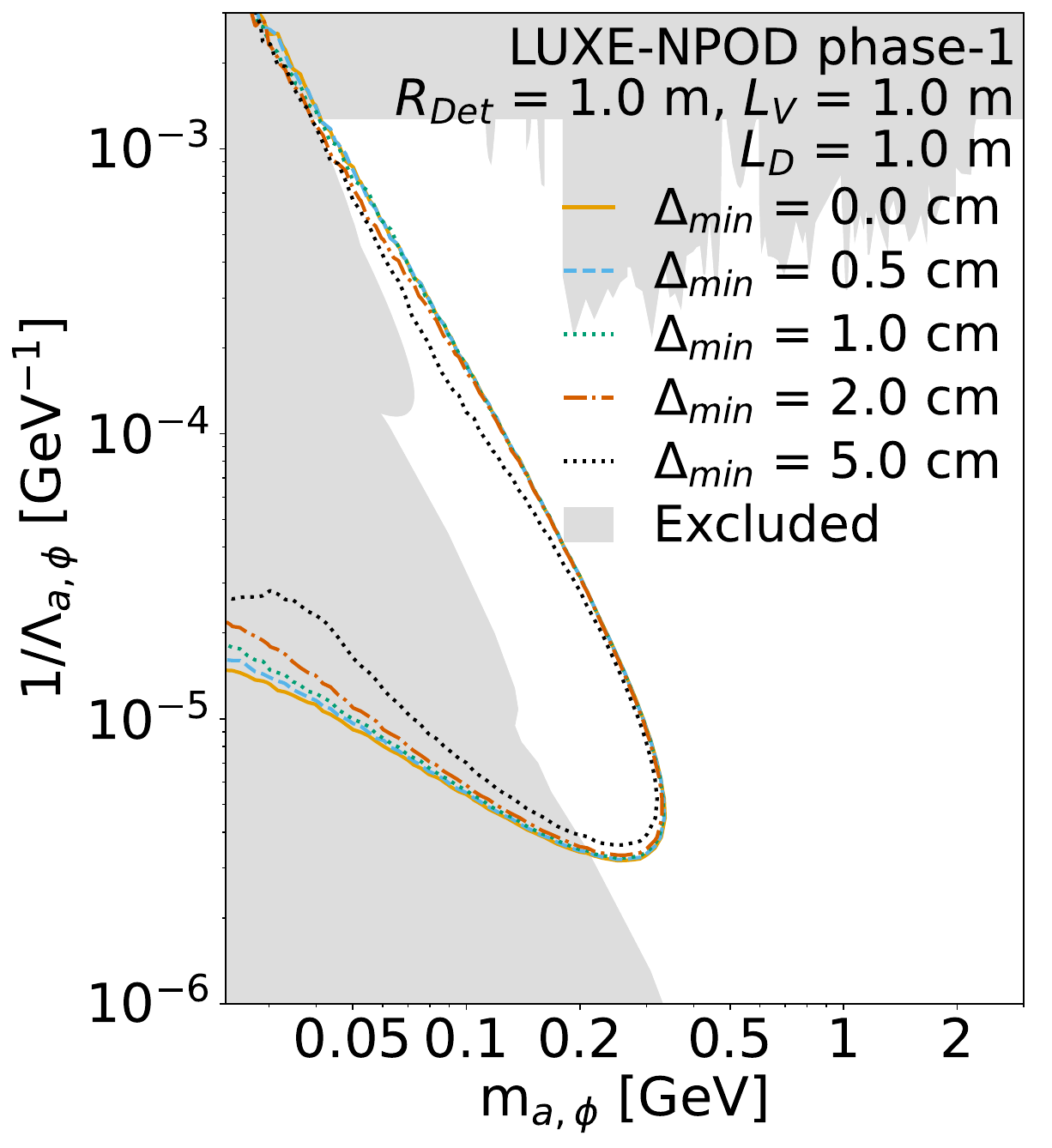}
        \caption{}
        \label{subfig:phase1_sensitivity_minseparation}
    \end{subfigure}
    \caption{Sensitivity projections for LUXE-NPOD phase-1 in the plane of the coupling to photon versus ALP mass. The sensitivity is inspected in terms of different dump length $L_D$ (a), decay volume length $L_V$ (b) and minimum di-photon separation at the detector $\Delta_\text{min}$ (c). For (a) and (b) there is no condition on $\Delta_\text{min}$. The exclusion limits indicate where more than three signal events are expected assuming no ALPs are observed, a background-free environment, and a ``virtual'' ideal detector with $1 \m$ radius for any photon above 0.5~GeV.
    The gray areas are currently existing bounds from other other experiments~\cite{PhysRevLett.125.081801, PhysRevLett.123.071801, PhysRevLett.118.171801, PhysRevLett.125.161806, Ablikim_2023, PhysRevD.110.L031101, BERGSMA1985458, PhysRevLett.59.755, Dolan_2017, na62_dump, PhysRevD.108.075019, faser_2025}.}
    \label{fig:sensitivity_phase1}
\end{figure}

We initially focus our feasibility study by maximizing signal efficiency, assuming LUXE phase-1 scenario and using different dump lengths ($L_D$), decay volumes ($L_V$), detector radii ($R_\text{Det}$), and di-photon separation at the detector ($\Delta_\text{min}$).
A selection on the photons energy is imposed at $E_{\gamma} > 0.5\GeV$, to emulate the soft backgrounds rejection requirements of the real experiment.  
This study is done initially with a ``virtual'' circular detector of 1~m radius as a benchmark, and a dump radius $~R_D~=~50~\cm$. 
The impact of smaller detector radii down to 0.1~m on the sensitivity degradation is negligible. 
The results are shown in Fig.~\ref{fig:sensitivity_phase1} and summarized below.

Fig.~\ref{subfig:phase1_sensitivity_dumplength} compares the impact of different dump lengths ($L_{D}$) assumptions. 
A short dump can improve signal efficiency for particles with shorter lifetimes, enhancing exclusion limits at higher masses, but at the cost of larger background. In contrast, a long dump can reduce the background, but also the signal with a relatively short lifetime, such that the ALP has a high probability to decay inside the dump.
To illustrate the sensitivity gain of shorter dumps, this study is done with a decay volume length $L_{V} = 2.5\m$, irrespective of the di-photon shower separation, to allow a direct comparison with the phase-1 results of~\cite{Bai_2022}.
The gain in sensitivity manifests as the boundaries of the inspected phase space are pushed to higher uncharted masses, for couplings in the $10^{-5} - 10^{-3} \GeV^{-1}$ range. 
However, this gain in sensitivity comes at a cost of an increased background, 
which makes the search more challenging in terms of the requirements on the detector. 
This is discussed in more detail in Sec.~\ref{sec:background}

Likewise, Fig.~\ref{subfig:phase1_sensitivity_decayvolume} compares the impact of different decay volumes ($L_V$) assumptions.
A long decay volume favors the sensitivity to long ALP lifetime, i.e., small ALP particle couplings.
Additionally, it allows the decay products of the ALPs to separate in the transverse plane before reaching the detector, making it easier to resolve them.
On the other hand, this also implies that a larger detector surface is needed to ensure a high signal acceptance.
We assume a dump length $L_{D} = 1.0\m$, irrespective of the di-photon shower separation.
We note that small couplings for small masses are already probed or will be probed by other beam dump experiments.
In addition, this region of the parameter space is already excluded by naturalness (see~\cite{Bai_2022}). Therefore, LUXE-NPOD does not effectively cover a new parameter space by using longer decay volumes.

Finally, Fig.~\ref{subfig:phase1_sensitivity_minseparation} compares the impact of different possible di-photon shower separation ($\Delta_\text{min}$).
This is an important parameter to take into account, since it directly affects the ALP sensitivity via the ability of the detector to resolve close-by-photon showers.
It is particularly relevant in case of short decay volumes that do not allow the photons to separate before reaching the detector.
To cope with (i) the typically limited space available for the entire experimental setup, and (ii) the fact that the decay volume medium is air, we select the shortest, but still effective decay volume of $L_V=1\m$.
We then analyze different values of photon shower resolution, $\Delta_{\rm min}$ in the $0-5\cm$ range.
The unrealistic case of $\Delta_{\rm min}=0~\cm$ is only shown for reference.
It can be seen that detectors capable of resolving photon showers separated by $\sim 2\cm$ or more provide the same results as an ideal detector in the region of interest for LUXE-NPOD.
This aspect is discussed further in Sec.~\ref{sec:detector}.

To sum up the signal discussion, LUXE-NPOD can in principle afford a short decay volume of $1\m$ and a small detector, provided that it is able to resolve the two photons showers.

\section{Background estimation}
\label{sec:background}
As discussed in Ref.~\cite{Bai_2022} and confirmed in this study, the expected SM background reaching the detector is mainly made of neutrons and photons, which can mimic the di-photon signal at the detector.
The background coming from charged particles (mostly protons and pions) is not as significant as (i) it comes in lower numbers and (ii) the particles are of very low energy. 
The residual charged particles can be bent away by adding in the decay volume a relatively moderate magnet in terms of active length and field strength and hence this component is not studied further in this work.
\begin{figure}[!t]
\centering
    \begin{subfigure}[b]{0.74\textwidth}
        \centering\includegraphics[width=\textwidth]{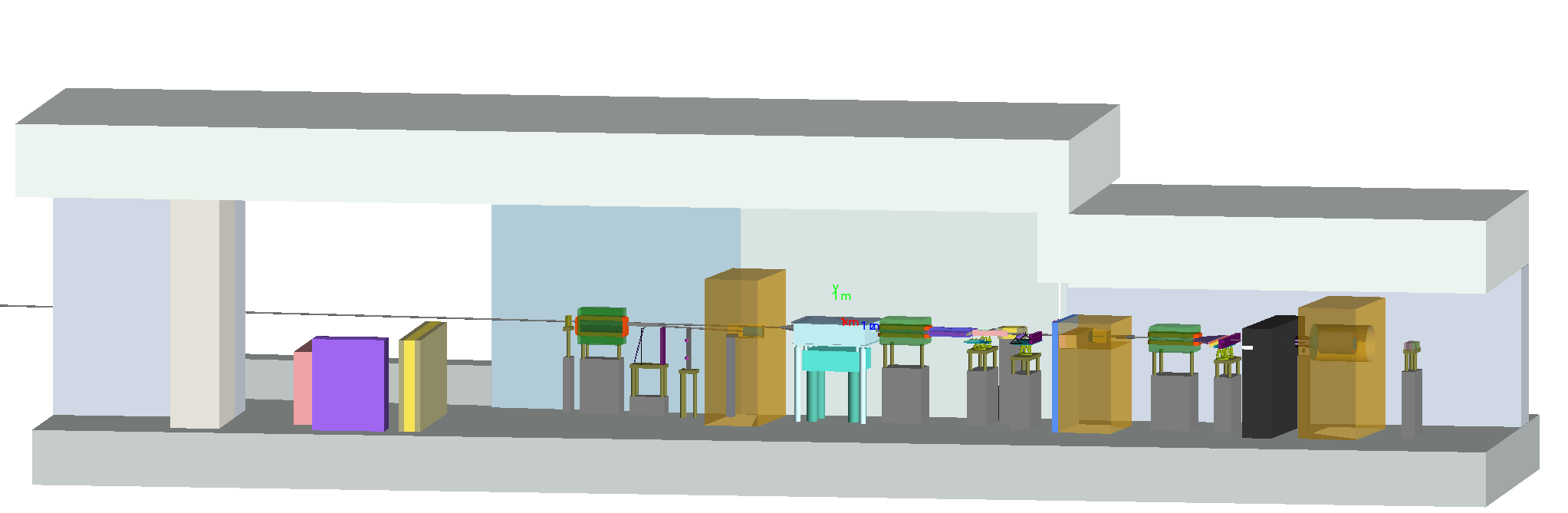}
        \caption{}
        \label{subfig:g4_luxe}
    \end{subfigure}
\hfill
    \begin{subfigure}[b]{0.24\textwidth}
        \centering\includegraphics[width=\textwidth]{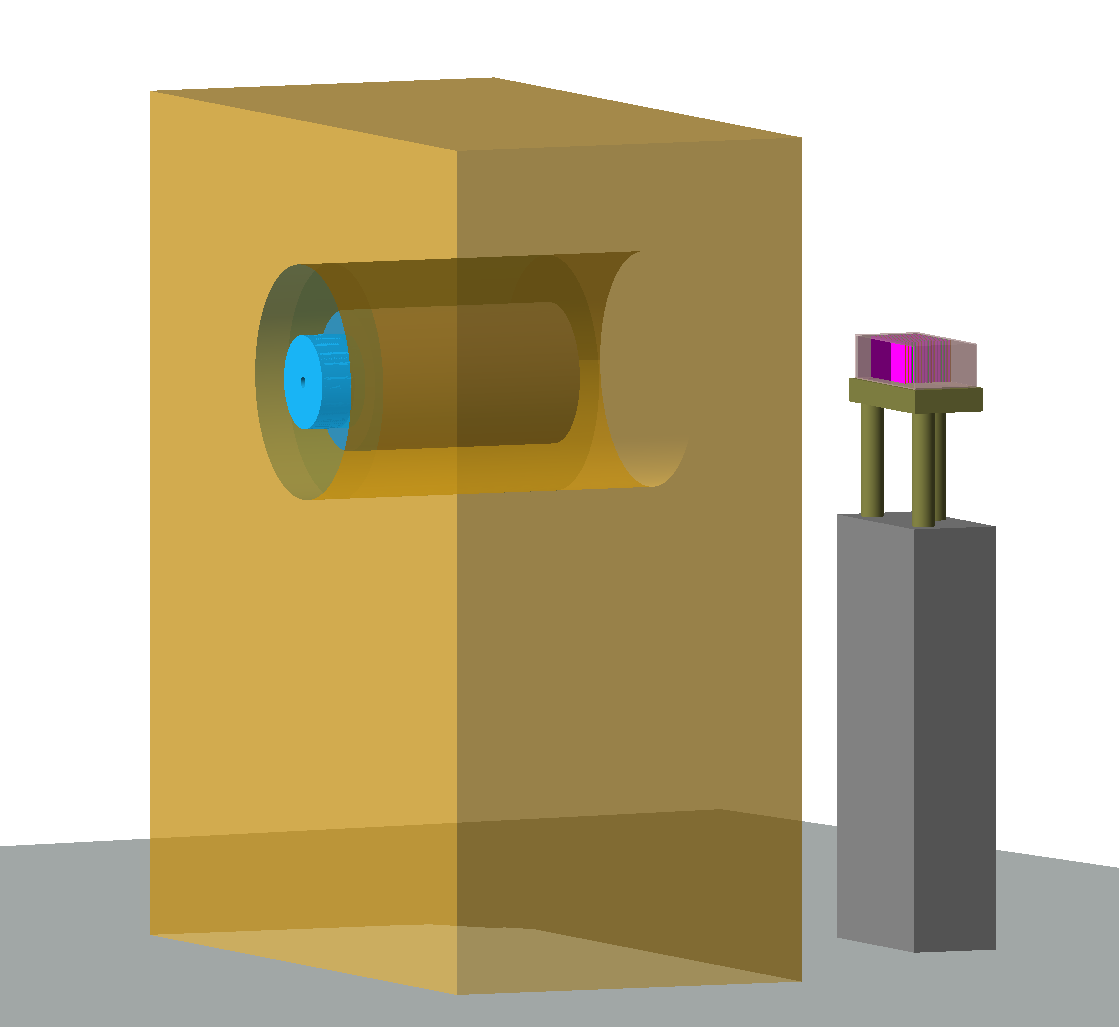}
        \caption{}
        \label{subfig:g4_bsm}
    \end{subfigure}
\caption{Visualization of the full LUXE \GEANTfour model for the $e$-laser setup~(a), and and a enlarged version of the same model, focused on the concrete wall (yellow) with the embedded dump in front of the photon detector~(b). The tunnel walls are removed from the model for visualization purpose.}
\label{fig:backgrounds_evt_disp_1}
\end{figure}

To estimate the background, we simulate the full LUXE apparatus as seen in Fig.~\ref{fig:backgrounds_evt_disp_1}. 
Unlike the preliminary study in~\cite{Bai_2022}, which used a very simplified model with only the dump in vacuum, our updated approach allows for a reliable estimation of back-scattering of particles with any element of the LUXE apparatus and its experimental hall based on~\cite{abramowicz2023technical}.

The \PTARMIGAN NCS photons output is propagated by \GEANTfour~v11.0.3~\cite{AGOSTINELLI2003250,ALLISON2016186,geant1610988} from the $e$-laser interaction point (IP) towards the dump, which is at a distance of $13.7 \m$.
The interaction of these NCS photons with the dump material and the evolution of the shower is simulated using the physics list \texttt{QGSP\_BERT\_HP} of \GEANTfour, which includes an updated neutron model compared to the one used in~\cite{Bai_2022}.
Finally, the original study assumed a pencil-like NCS photon beam and only the energy spectrum was properly simulated, while here the full NCS beam profile and divergence is simulated, starting from the electron beam and the polarized laser, by \PTARMIGAN and then propagated by \GEANTfour.

To stay conservative, we perform the optimization study for the dump design using the incoming phase-1 NCS photon spectra, which naturally leads to the largest background among the three phases.
We first compare the same tungsten dump size as in~\cite{Bai_2022}: dump length $L_{D} = 1\m$, dump radius $R_{D} =50\cm$, detector radius $R_\text{Det} = 1\m$, decay volume $L_{V} = 2.5\m$, and physics list \texttt{QGSP\_BERT}.
Compared to the simplified model discussed in~\cite{Bai_2022}, the background expectations are confirmed by the observed distributions of photons and neutrons.
Nevertheless, a significant drawback lies in the high cost associated with a tungsten dump of such magnitude. Consequently, a comprehensive study is conducted across various configurations to optimize the dump design, aiming to reduce tungsten usage and associated costs without compromising signal acceptance or background rejection performance.
Table~\ref{table:phase1optimizationbkg} shows the number of background photons ($N_\gamma$) and neutrons ($N_n$) per BX, when simulating different dump materials and sizes.
Simulating many BXs is computationally expensive and we therefore begin the optimization study by simulating 50\% of a full bunch crossing for each configuration.
We begin with a tungsten dump of length $L_{D} = 30\cm$ and radius $R_{D} =10\cm$ (Table~\ref{table:phase1optimizationbkg}.a), and as expected we see a significant background increase.
We investigate the origin of each particle by tracking the trajectory they follow to reach the detector.
We observe that the ideal case, in which particles enter the dump and travel in the positive $z$ direction towards the detector, agrees with the previous results in~\cite{Bai_2022}. 
The increase rather originates from particles escaping through the sides of the dump and multiple scattering, from the dump back to other elements of the experimental setup, and finally to the detector.
While the latter case produces very soft particles, with energies well below $0.1\GeV$, particles escaping from the side of the dump can reach the detector with energies above the $0.5\GeV$ threshold.
Thus, a small-radius dump alone is not enough to suppress the background.
Since we see that most of the background comes from particles escaping from the side of the dump, we add a concrete wall enclosure to the design of the dump (Fig.~\ref{subfig:g4_bsm}).
Even if we increase the dump length $L_D$ (Table~\ref{table:phase1optimizationbkg}.b and Table~\ref{table:phase1optimizationbkg}.c) for a small dump radius $R_D$, a significant amount of background still persists above $0.5 \GeV$ for the short dump variations compared to the $1\m$ nominal case.
By analyzing the trajectory of the neutrons above $E_\text{kin}=0.5 \GeV$, we see that most of them are originating from the beginning of the dump and escaping through the sides of the concrete wall.
Hence, we increase the dump radius $R_D$ (Table~\ref{table:phase1optimizationbkg}.d and Table~\ref{table:phase1optimizationbkg}.e), where we see a significant decrease in neutrons population above $E_\text{kin}=0.5 \GeV$.
As tungsten is expensive and difficult to handle, we also test other materials with a larger $R_D$.
A dump made completely of lead (Table~\ref{table:phase1optimizationbkg}.f) does not provide a competitive stopping power, while also reducing the signal production rate due to its larger radiation length by a factor $\sim 0.6$. 
Since the NP production happens within the first few mm's of the dump, we add another configuration with a core of tungsten of radius $20 \cm$, wrapped in a hollow lead cylinder with an inner radius of $20 \cm$ and an external radius of $50 \cm$, where both the core and wrap are of length $L_D = 1 \m$ (Table~\ref{table:phase1optimizationbkg}.g).
This design effectively suppresses the background from photons and neutrons with $E_\text{kin}~>~0.5 \GeV$.

\begin{table}[!ht]
\small\setlength\tabcolsep{4.5pt}
\begin{tabular}{cccc|cc|cc}
\hline
  Setup, $N_{\rm BX}$ &
  \begin{tabular}[c]{@{}c@{}}Dump\\ material\end{tabular} &
  $L_D$ [cm] &
  $R_D$ [cm] &
  \begin{tabular}[c]{@{}c@{}}$N_\gamma$ \\ all \end{tabular} &
  \begin{tabular}[c]{@{}c@{}}$N_\gamma$ \\ $E_\text{kin} > 0.5 \GeV$\end{tabular} &
  \begin{tabular}[c]{@{}c@{}}$N_n$ \\ all \end{tabular} &
  \begin{tabular}[c]{@{}c@{}}$N_n$ \\ $E_\text{kin} > 0.5 \GeV$\end{tabular} \\
\hline
a, 0.5 & \multirow{5}{*}{W} & 30  & 10  & 9500  & 60 & 970000  & 26000  \\
b, 0.5 &  & 50  & 10  & 2600 & 120 & 320000 & 5800 \\
c, 0.5 &  & 100 & 10  & 3300 & 0 & 140000 & 1600 \\
d, 0.5 &  & 100 & 20  & 840 & 0 & 39000  & 210  \\
e, 0.5 &  & 100 & 30  & 300 & 0 & 14000  & 75  \\
\hline
f, 0.5 & Pb  & 100 & 40  & 900 & 0 & 84000 & 1900 \\
\hline
g, 0.5 & \multirow{2}{*}{W+Pb} & \multirow{2}{*}{100} & W: 20 & 210 & 0 & 13000  & 30 \\
g, 10 &  &  & Pb: 50 & 160 & 0 & 12000 & 68 \\
\end{tabular}
\caption{Number of background photons ($N_\gamma$) and neutrons ($N_n$) arriving at the detector per BX for LUXE-NPOD phase-1.
The dump is enclosed in a concrete wall with an effective decay volume of $L_V=1\m$, and ``virtual'' detector of radius $1\m$.
Different dump materials and sizes are tested. The W+Pb refers to a core tungsten dump of $R=20\cm$, and lead wrap of $R_{\rm min}=20\cm$ and $R_{\rm max}=50\cm$. Each setup is simulated for a number $N_{\rm BX}$ of BX. $N_\gamma$ and $N_n$ are normalized to 1 BX. 
}
\label{table:phase1optimizationbkg}
\end{table}

\begin{figure}[htbp]
    \centering
    \begin{subfigure}[b]{0.32\textwidth}
        \centering
        \includegraphics[width=\textwidth]{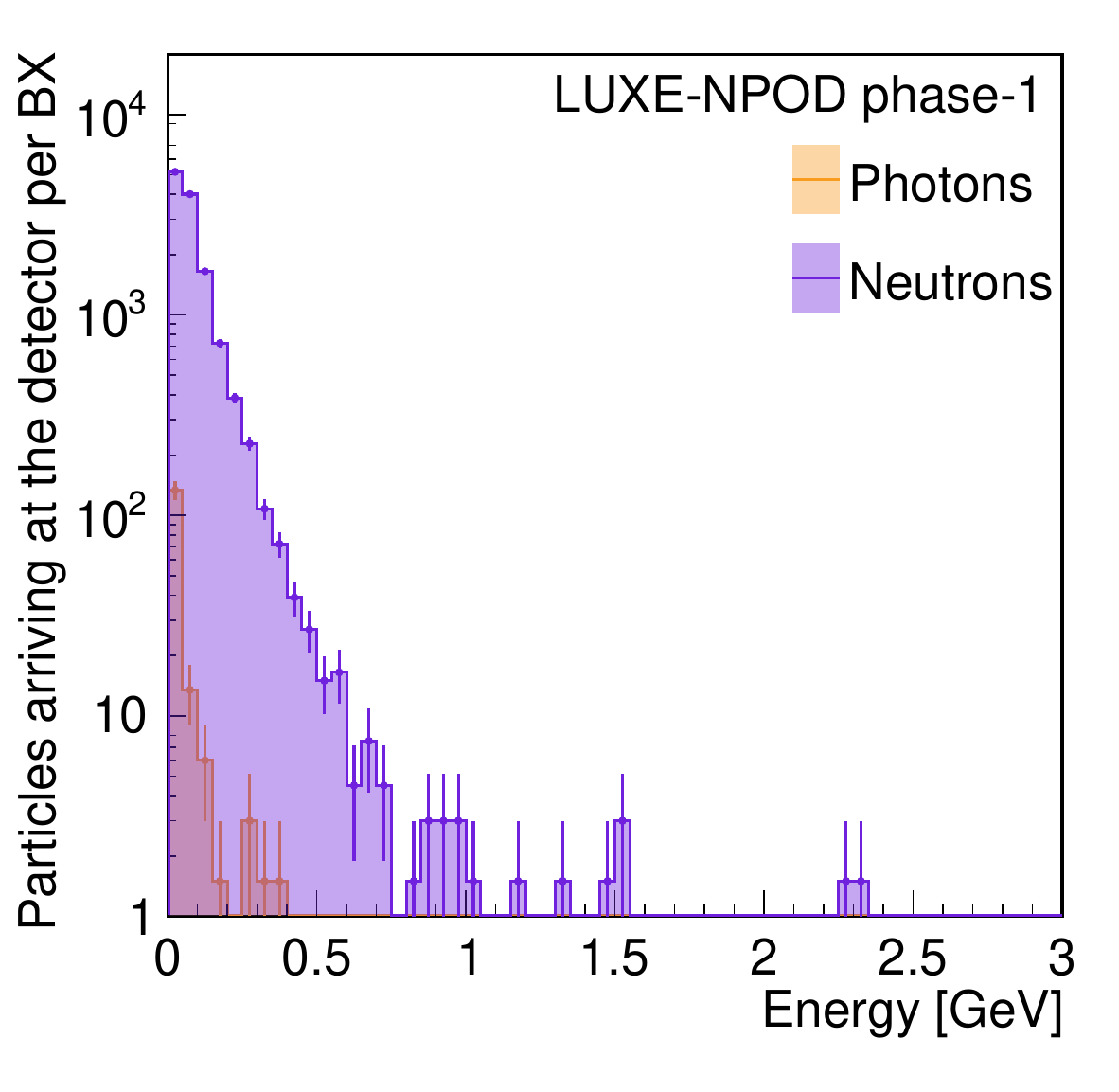}
        \caption{}
        \label{subfig:doubledump_energy}
    \end{subfigure}
    \hfill
    \begin{subfigure}[b]{0.32\textwidth}
        \centering
        \includegraphics[width=\textwidth]{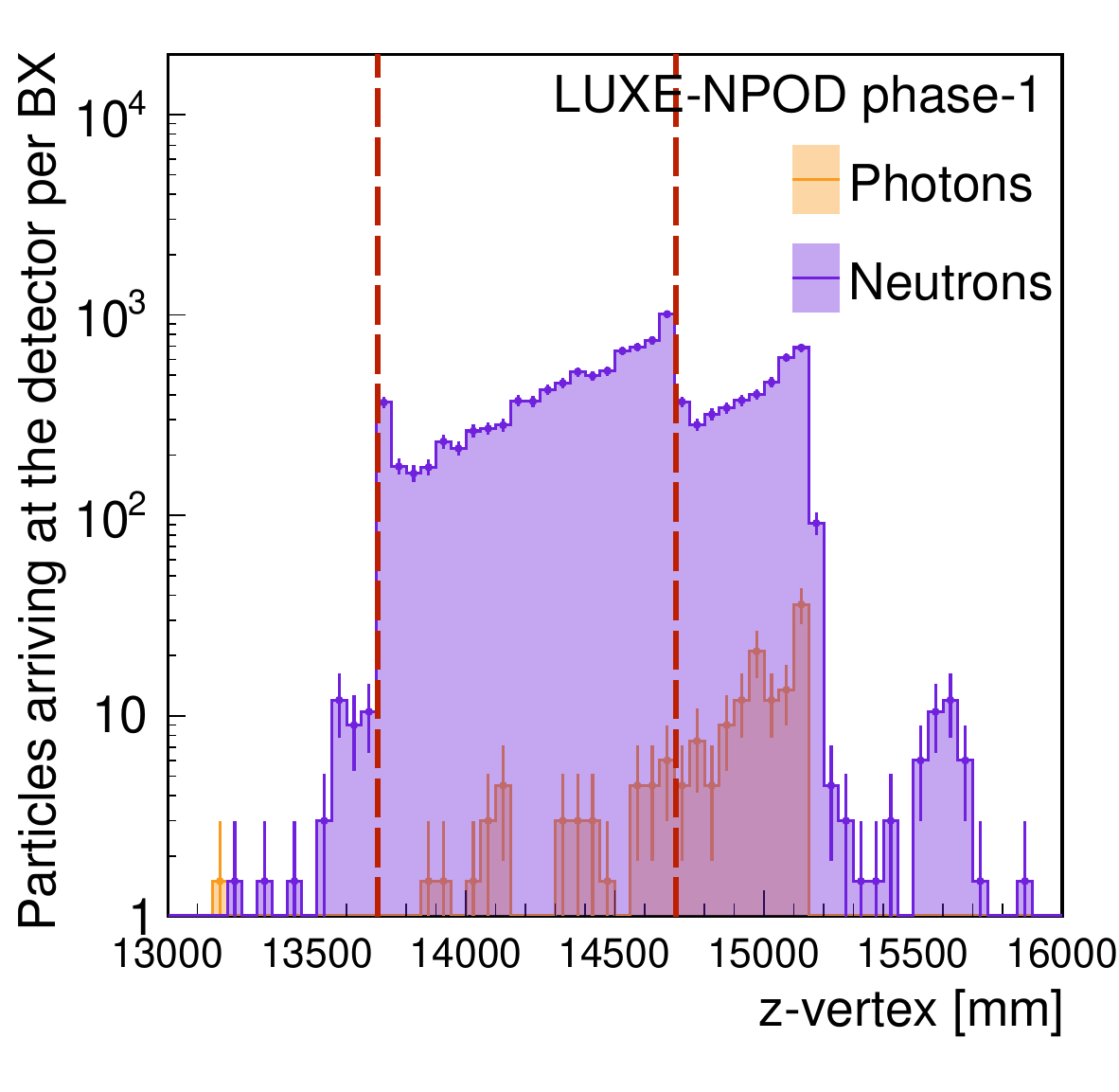}
        \caption{}
        \label{subfig:doubledump_vtxz}
    \end{subfigure}
    \begin{subfigure}[b]{0.32\textwidth}
        \centering
        \includegraphics[width=\textwidth]{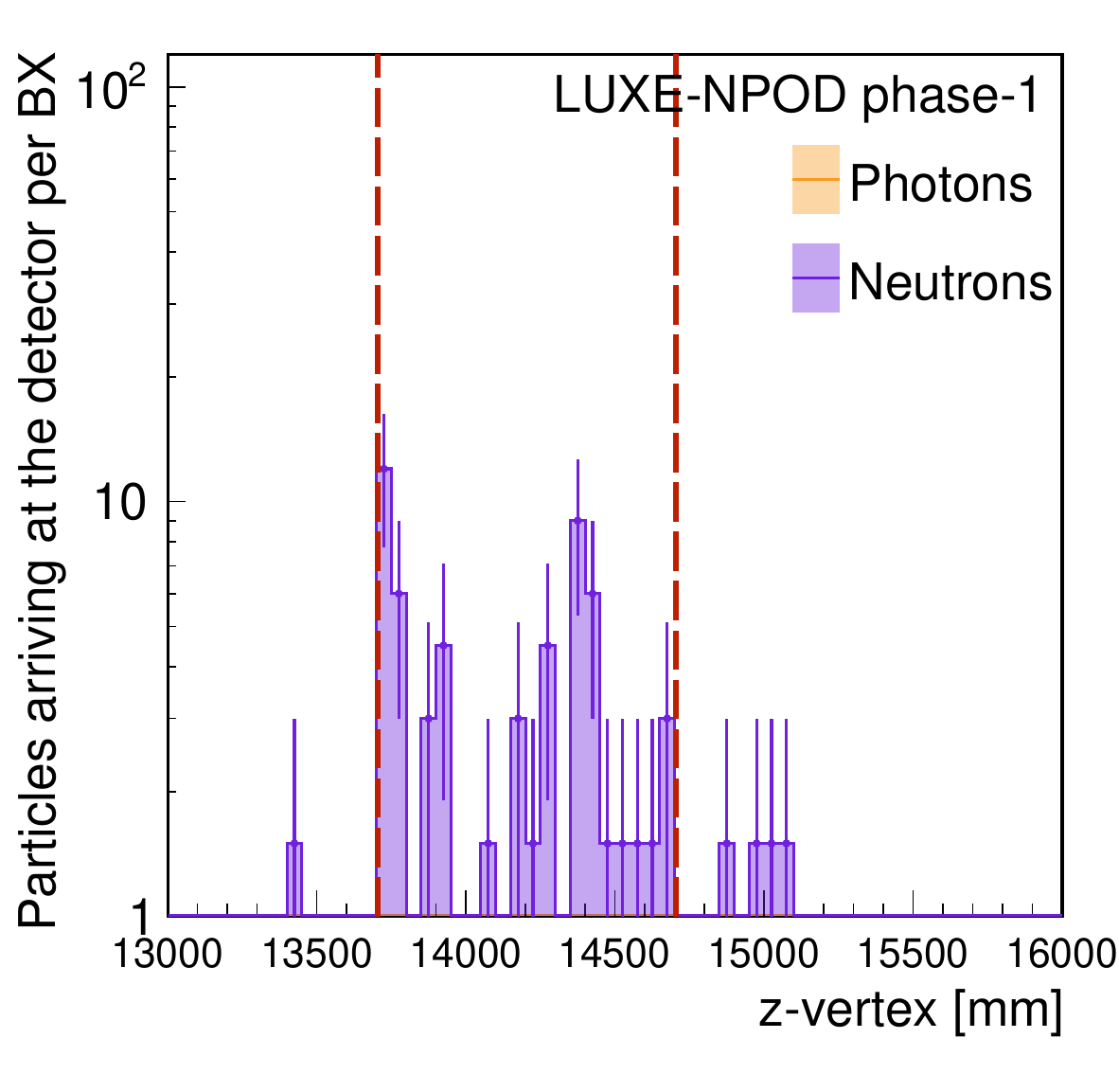}
        \caption{}
        \label{subfig:doubledump_vtxz_ecut}
    \end{subfigure}
    \begin{subfigure}[b]{0.32\textwidth}
         \centering
         \includegraphics[width=\textwidth]{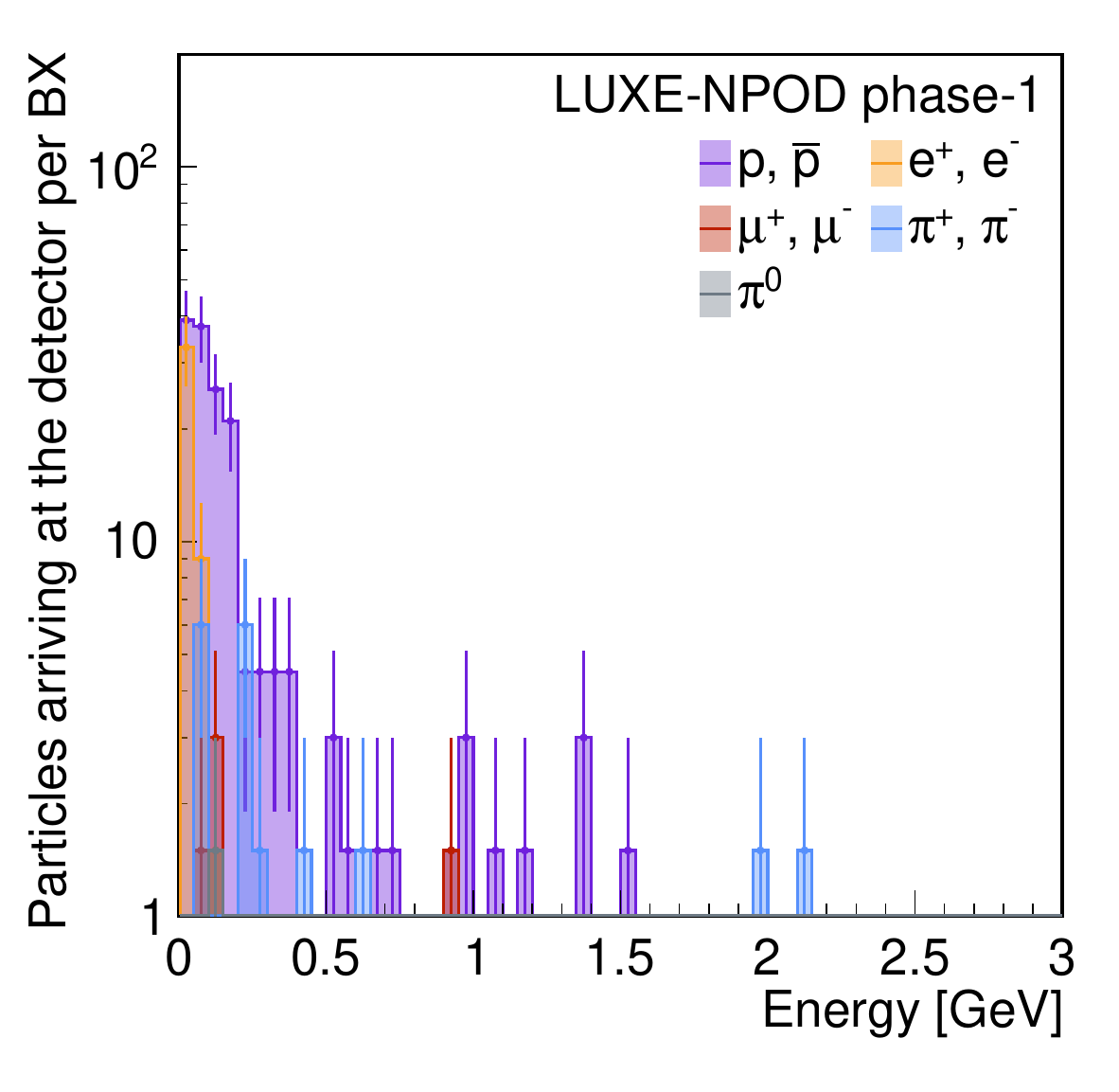}
         \caption{}
         \label{subfig:charged_energy}
    \end{subfigure}
    \hfill
    \begin{subfigure}[b]{0.32\textwidth}
         \centering
         \includegraphics[width=\textwidth]{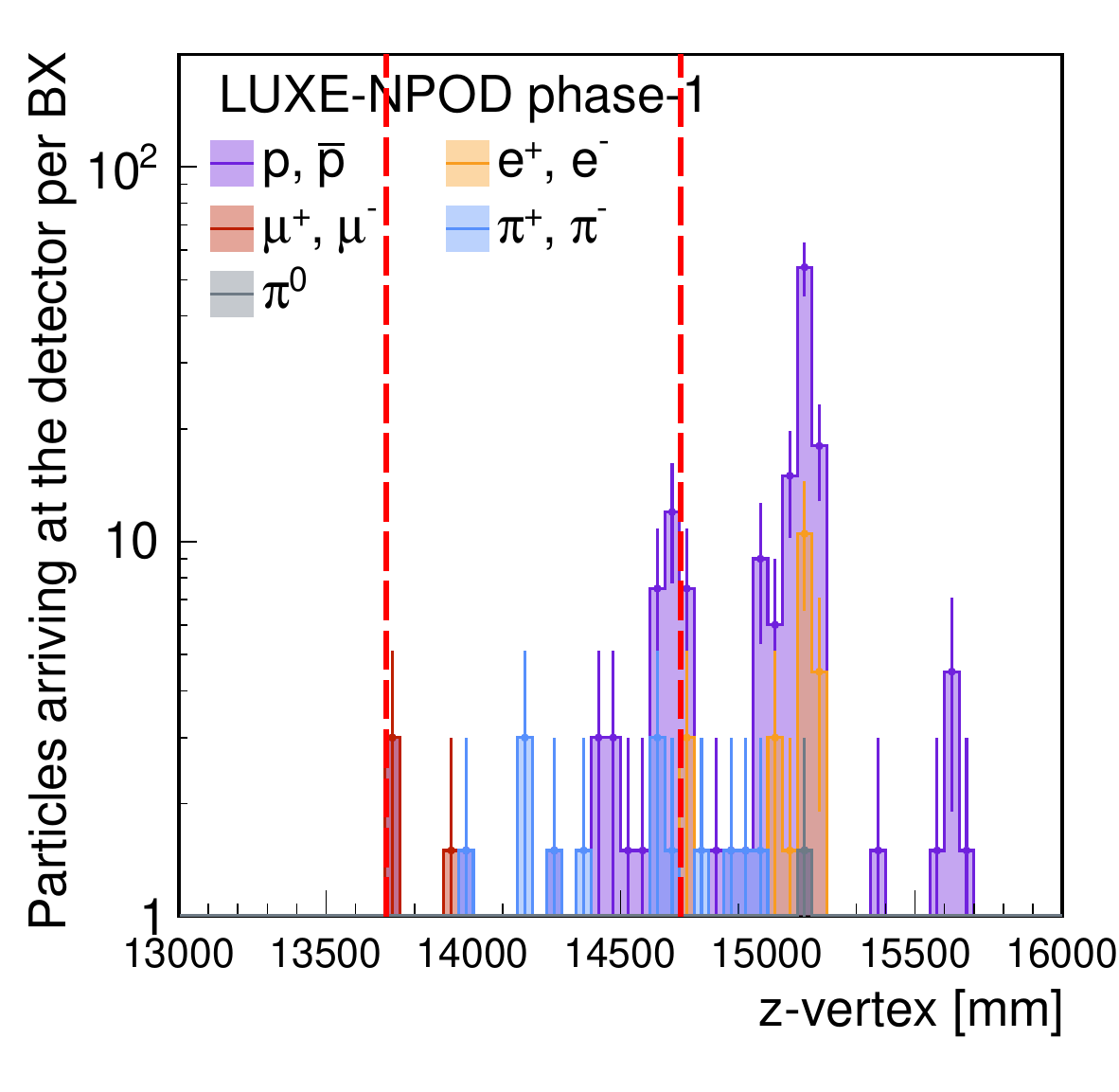}
         \caption{}
         \label{subfig:charged_vtxz}
    \end{subfigure}
    \hfill
    \begin{subfigure}[b]{0.32\textwidth}
         \centering
         \includegraphics[width=\textwidth]{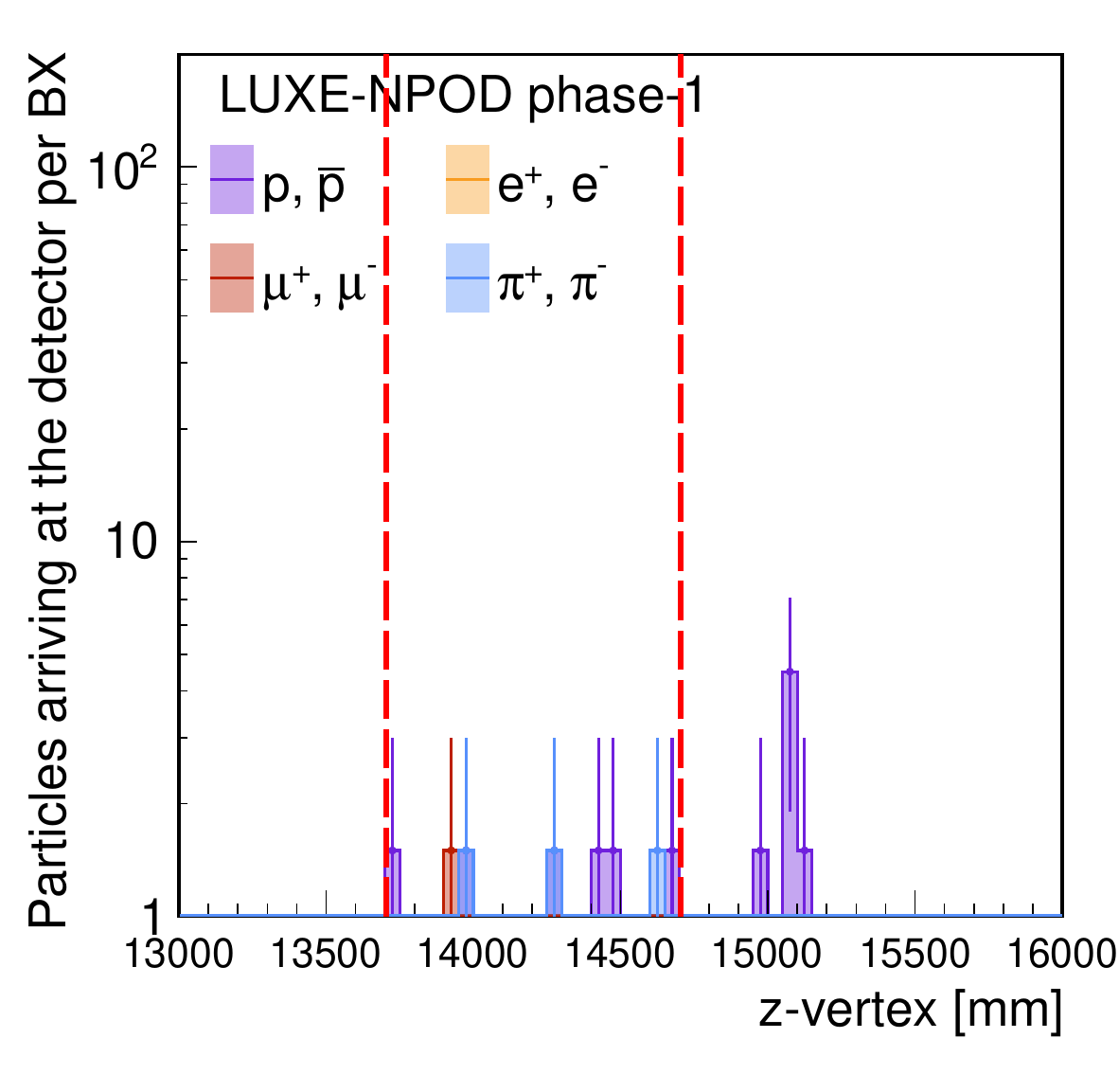}
         \caption{}
         \label{subfig:charged_vtxz_ecut}
    \end{subfigure}    
    \caption{Photons and neutrons energy (a), $z$-vertex (b) and $z$-vertex with $E_\text{kin} > 0.5 \GeV$ (c) distributions. Long-lived neutral and charged particles energy (d), $z$-vertex (e) and $z$-vertex with $E_\text{kin} > 0.5 \GeV$ (f) distributions. The design used is a core tungsten dump of $R_D=20 \cm$, $L_D=100 \cm$, and lead wrap of $R_D=50 \cm$. The dump is enclosed in concrete, the decay volume is $1 \m$, and the detector has a radius of 1 m. The beginning and end of the dump are depicted with vertical dashed lines in (b), (c), (e) and (f). The simulation is done for 10 BXs.}
    \label{fig:doubledump_E_vtxz}
\end{figure}

Following this coarse scan, and to obtain a reliable estimation of $N_\gamma$ and $N_n$, we simulate 10 full BXs using the W+Pb design (Table~\ref{table:phase1optimizationbkg}.g, 10).
We remind that the analysis in~\cite{Bai_2022} used only two BXs.
Figs~\ref{subfig:doubledump_energy},~\ref{subfig:doubledump_vtxz} and~\ref{subfig:doubledump_vtxz_ecut} show the energy and the production vertex $z$ coordinate distributions for photons and neutrons (that reach the detector) respectively.
While no photons are seen with $E_\text{kin}~>~0.5 \GeV$, 68 neutrons per BX with $E_\text{kin}~>~0.5 \GeV$ arrive at the detector. \\

\begin{figure}[!ht]
    \centering
    \begin{subfigure}[b]{0.495\textwidth}
        \centering
        \includegraphics[width=\textwidth]{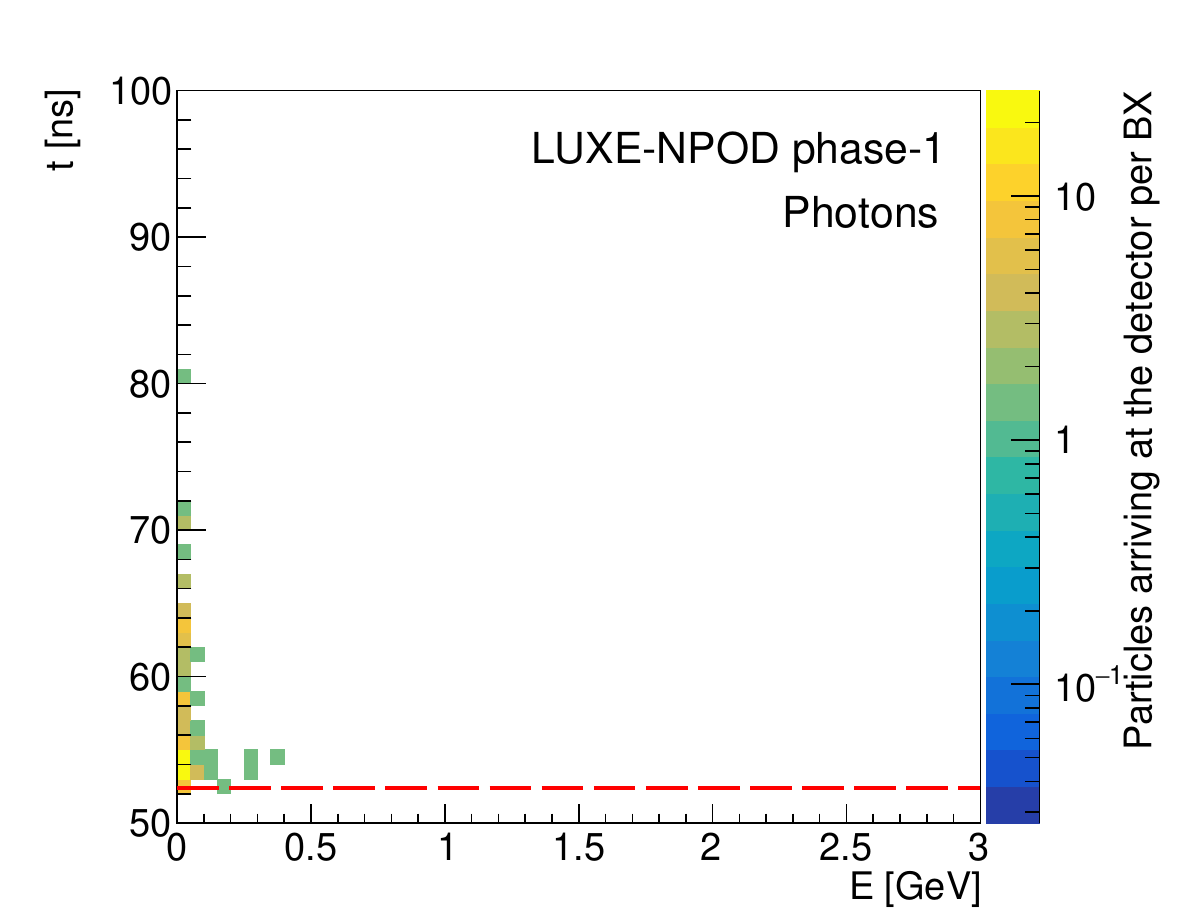}
        \caption{}
        \label{subfig:background_photons_time_of_arrival}
    \end{subfigure}
    \hfill
    \begin{subfigure}[b]{0.495\textwidth}
        \centering
        \includegraphics[width=\textwidth]{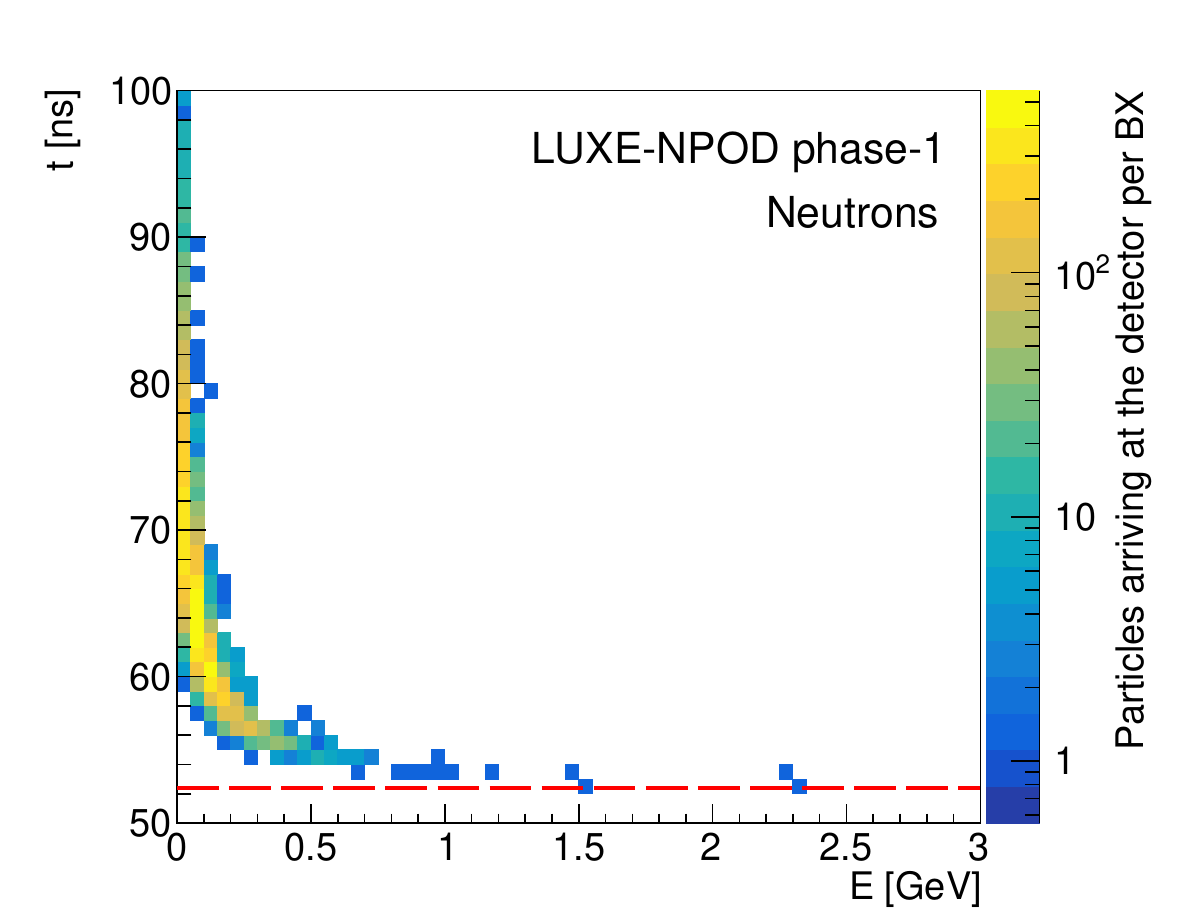}
        \caption{}
        \label{subfig:neutrons_time_of_arrival}
    \end{subfigure}
    \caption{Time of arrival at the detector of background photons (a) and background neutrons (b). The simulation considers 10 BXs for a W+Pb dump of length $L_\text{D} = 1\m$, decay volume of length $1 \m$, and detector radius of $1 \m$. The time of arrival is measured with respect to the moment of the $e$-laser interaction, when $t=0$. The horizontal dashed line shows the estimated time of arrival of the signal photons, at $t=52.4\ns$.}
    \label{fig:bkg_times_of_arrival}
\end{figure}

As most of the neutrons come at much lower energies as can be seen in Fig.~\ref{subfig:doubledump_energy}, they may arrive at the detector delayed in time compared to the signal photons.
Fig.~\ref{fig:bkg_times_of_arrival} shows the background photons and neutrons time of arrival measured from the e-laser interaction point versus their energies, while the signal photon's time of arrival is around $52.4\ns$, shown by the dashed line.
It is seen that neutrons are typically slightly delayed with respect to that, and a time resolution of about 0.1 to $1\ns$ may help identifying neutrons with energies up to about $1.5\GeV$.

With 10 BXs there is enough statistics to estimate the number of charged or long-lived neutral SM particles.
The expected background from charged particles (without any dedicated magnet to deflect them) is presented in Figs~\ref{subfig:charged_energy},~\ref{subfig:charged_vtxz} and~\ref{subfig:charged_vtxz_ecut}.
The rate of charged particles arriving at the detector at a minimum energy of $0.5\GeV$ is of 23 particles per BX.
As mentioned earlier, these particles can be bent away by placing a moderate magnet (field strength and length) right after the dump.
While we leave the feasibility question of adding such a magnet to a future work, we note that the charged component is a factor 3 smaller than the neutron one and hence it does not change the overall picture of the background, even without considering an additional magnet.

Finally, aiming for a shorter dump design, we also explore the possibility of implementing a magnetized dump, a solution that was proposed, e.g., in~\cite{VANELP1997403,March:1971ps}. 
By magnetizing the dump, the magnetic field can deflect charged particles (such as electrons, protons, and pions) within the electromagnetic and hadronic showers initiated by the incoming photon beam. As a result, neutrons will be produced at larger angles relative to the beam axis and, in a simplified view, are unlikely to reach the detector. Nevertheless, we find no significant gain in background suppression by adding a magnetic field to the dump and hence we choose the W+Pb dump configuration (Table~\ref{table:phase1optimizationbkg}.g) as the most efficient for LUXE-NPOD phase-1. Details about the magnetized dump study can be found in App.~\ref{app:magnetized_dump}

As we note in Sec.~\ref{sec:npodsensitivity}, the generic $1\m$ radius detector is somewhat redundant in terms of the acceptance and hence the background for a smaller detector will be linearly smaller.
In the following section, we discuss a specific detector choice, which indeed has a smaller size than $1\m$ radius.

\section{Detector performance}
\label{sec:detector}
As quantified in the discussions above, the detector for LUXE-NPOD faces several challenges to enhance the signal efficiency and to effectively reject the residual background that may emerge from the dump.
To ensure high signal efficiency, the detector must have a sufficiently large geometric acceptance and the ability to resolve nearby photon showers.
The studies in Sec.~\ref{sec:npodsensitivity} conclude that a detector of few tens of centimeters in diameter of active area will be enough to cover the phase-space of interest.
These results also indicate that di-photon showers, which can be separated by $\sim 2\cm$ or more at the surface of the detector lead to a good signal efficiency.
That is, the efficiency loss due to more collimated di-photon systems will not harm the sensitivity significantly.
Regarding the effective rejection of the residual background, as discussed in Sec.~\ref{sec:background}, hard neutrons may represent a background if they mimic the signature of photons in the detector.

\begin{figure}[!ht]
    \centering
    \begin{subfigure}[b]{0.55\textwidth}
        \centering
        \includegraphics[width=\textwidth]{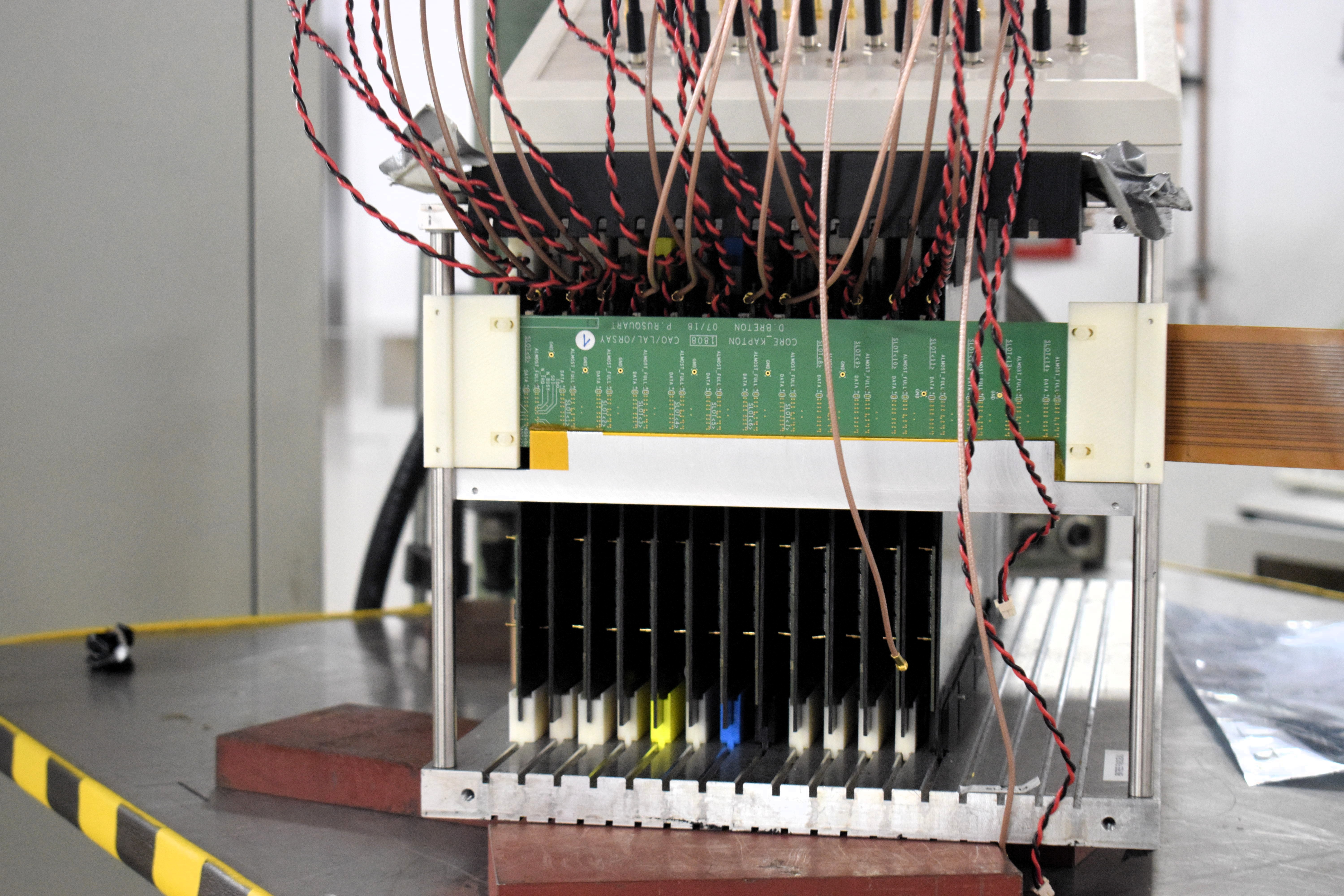}\\
        \caption{}
        \label{subfig:calice_prototype}
    \end{subfigure} \\
    \vfill
    \begin{subfigure}[b]{0.475\textwidth}
        \centering
        \includegraphics[width=\textwidth]{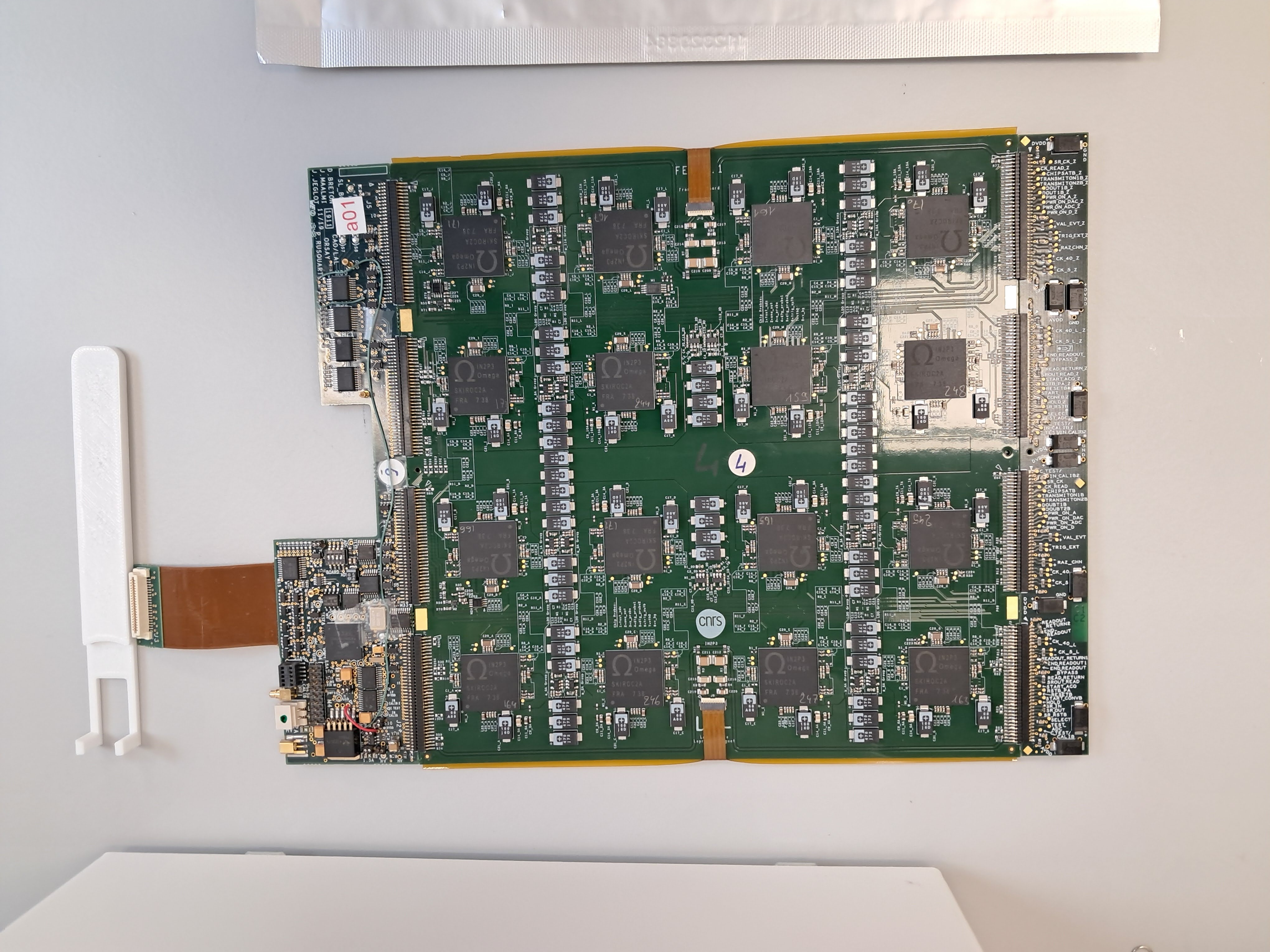}
        \caption{}
        \label{subfig:calice_modules}
    \end{subfigure}
    \hfill
    \begin{subfigure}[b]{0.475\textwidth}
        \centering
        \includegraphics[width=\textwidth]{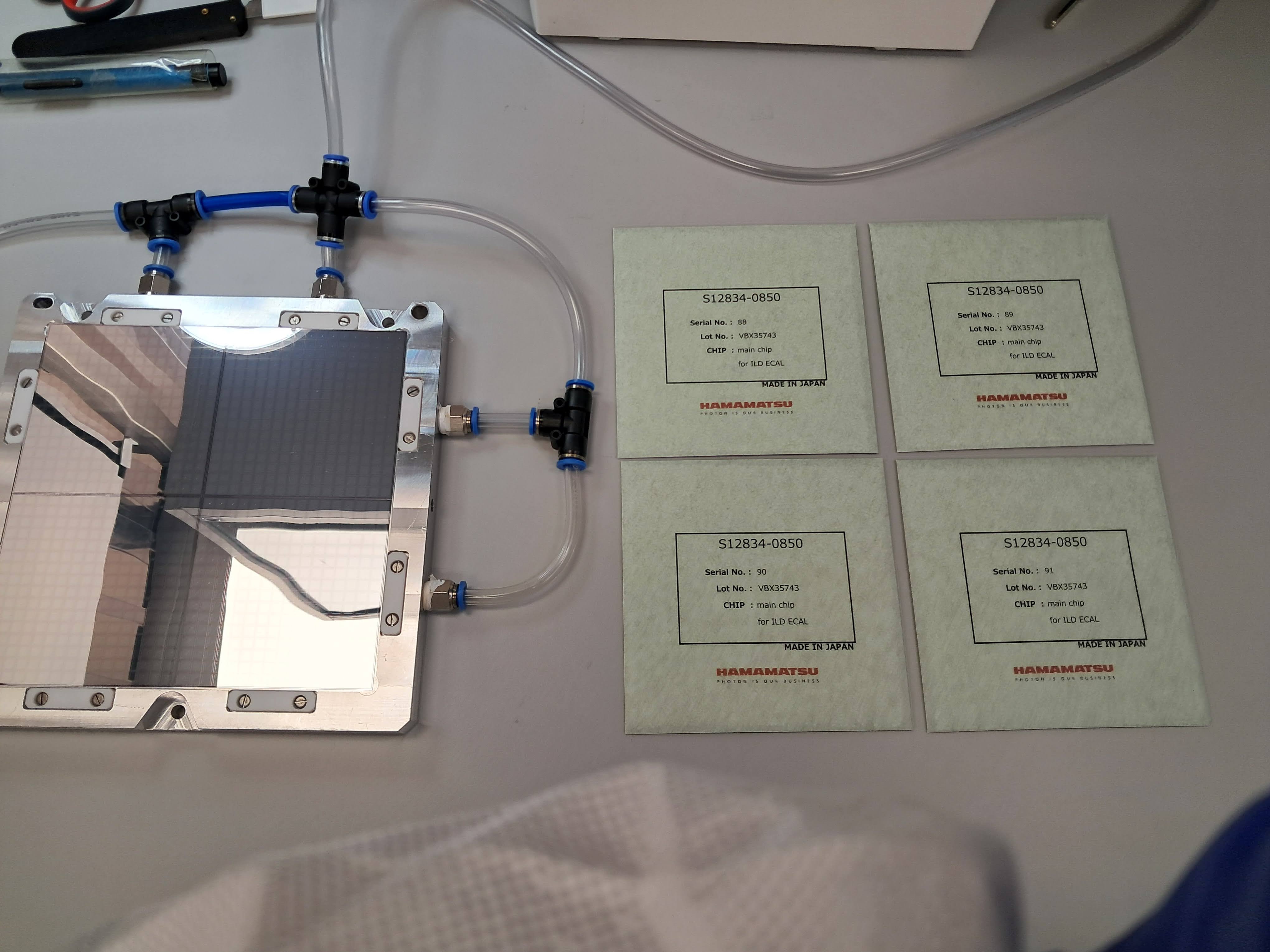}
        \caption{}
        \label{subfig:calice_sensors}
    \end{subfigure}
    \caption{Picture of several layers of the CALICE SiW-ECAL prototype from 2021 (a). Different configurations of tungsten repartition and layer sorting were tested, with a maximum of 15 active layers tested simultaneously and a total depth of up to $24\xzero$~\cite{Okugawa:2024dks}. 
    All layers consist in single modules of $18\times18$ cm$^{2}$ (b) equipped with 4 sensors each (c).}
    \label{fig:detector_cad}
\end{figure}

These challenges can be addressed using the electromagnetic calorimeter for the electron side (ECAL-E) in the $\gamma$-laser mode of LUXE.
The LUXE ECAL-E~\cite{abramowicz2023technical} is a sandwich silicon-tungsten electromagnetic calorimeter based on the SiW-ECAL~\cite{Kawagoe:2019dzh,Breton:2020xel}, designed and assembled by the CALICE collaboration~\cite{CALICE_TWIKI} and the DRD6 collaboration~\cite{CERN-DRDC-2024-004} for calorimetry (one of the new detector R\&D collaborations approved by CERN).
Its adapted design for LUXE foresees 15 layers equipped with tungsten repartition that ranges from $18 \xzero$ to up to $26 \xzero$.
Silicon sensors, featuring $5.53\times 5.53~{\rm mm}^2$ pads in $90\times90~{\rm mm}^2$ area, are distributed such that they will cover a surface of $180\times360~{\rm mm}^2$ and a total depth of up to 22.5~cm. 
The silicon wafer has a thickness of $500\mum$.
With this compactness and granularity, the LUXE ECAL-E ensures a small Moli\`ere radius that minimizes the transverse spread of electromagnetic showers, allowing tracking the longitudinal development of showers, and favoring the identification and rejection of the backgrounds.
With these characteristics, the LUXE ECAL-E could provide good single and double photon reconstruction and identify ALP decays.
Moreover, with a timing resolution of 0.5 to 1~ns provided by its ASIC readout SKIROC2A~\cite{Suehara:2018mqk}, this detector would also offer good capability for the rejection of slow neutron background contributions.
In this section, we present a performance study of this specific detector in conjunction with existing analysis algorithms.
Fig.~\ref{fig:detector_cad} shows a picture of the detector assembly prototype.

\subsection{LUXE ECAL NPOD simulation}
\label{subsec:detector_candidate}

For this study, we simulate the calorimeter for NPOD (LUXE ECAL NPOD) based on the ECAL-E design baseline.
The depth of the design is $22.5\cm$ that includes $18\xzero$ of tungsten.
In the design discussed here, each layer consists of one tungsten plate, $4.2\mm$ ($1.2\xzero$) in longitudinal depth.
The geometrical description of the detector is implemented using the DD4hep toolkit~\cite{Frank_2014, Gaede_2020}.
The simulation of the interactions between the isolated detector and the particles
are carried out with \GEANTfour~\cite{AGOSTINELLI2003250,ALLISON2016186,geant1610988} version \texttt{11.2.1} and
considering the \texttt{QGSP\_BERT\_HP} physics list.
Then, the simulation results are analyzed in the 
\texttt{Marlin} framework~\cite{Gaede_2006, Wendt_2007},
where several reconstruction algorithms are implemented, including a simple detector digitization, clustering, particle identification, tracking, and photon kinematic reconstruction.

Three sets of simulations are performed.
In the first set of simulations, pairs of photons with the same energy of $0.5\GeV$, $1.5\GeV$, and $3.5 \GeV$ are simulated, respectively, with normal incidence to the detector.
With these samples, we determine the efficiency for di-photon reconstruction, see Sec.~\ref{subsec:recon_clustering}.
In the second set, single photons of different energies and with different incident angles (with respect to the beam axis), are used to emulate one of the photons from the decay of an ALP.
With this sample, we evaluate the capabilities of topological reconstruction of the photon kinematics (entry point, angle of incidence), see Sec.~\ref{subsec:recon_vertex}.
For the last set, we shoot single particles straight towards the front surface of the detector, and the particle gun is set near the NCS photon beam axis, which coincides with the dump axis.
Both signal photon and potential background particles, such as neutron, proton, and pion, are simulated.
With these simulations, the background rejection capabilities due to the high-granularity of the LUXE ECAL NPOD are studied in Sec.~\ref{subsec:recon_pid}.

A scan over the photon energy in the range of $0.5 \GeV$ to $10 \GeV$ reveals the energy resolution of the detector as
\begin{equation}
    \dfrac{\sigma_E}{E} = \dfrac{(19.8 \pm 0.4) \%}{\sqrt{E_0/\GeV}} \oplus (4.9 \pm 0.3)\%,
    \label{eq:ecale_energy_resolution}
\end{equation}
where $E_0$ is the photon energy.
The resolution in Eq.~\eqref{eq:ecale_energy_resolution} includes the effects of the sensor pixelization and a simple readout digitization,
and is optimized for the said energy range. This estimation is consistent with the $\sim 14\%$ resolution measured at large energies and with a denser SiW-ECAL version~\cite{Okugawa:2024dks}.

\subsection{Shower separation}
\label{subsec:recon_clustering}

\begin{figure}[!ht]
    \centering
    \includegraphics[width=0.495\textwidth]{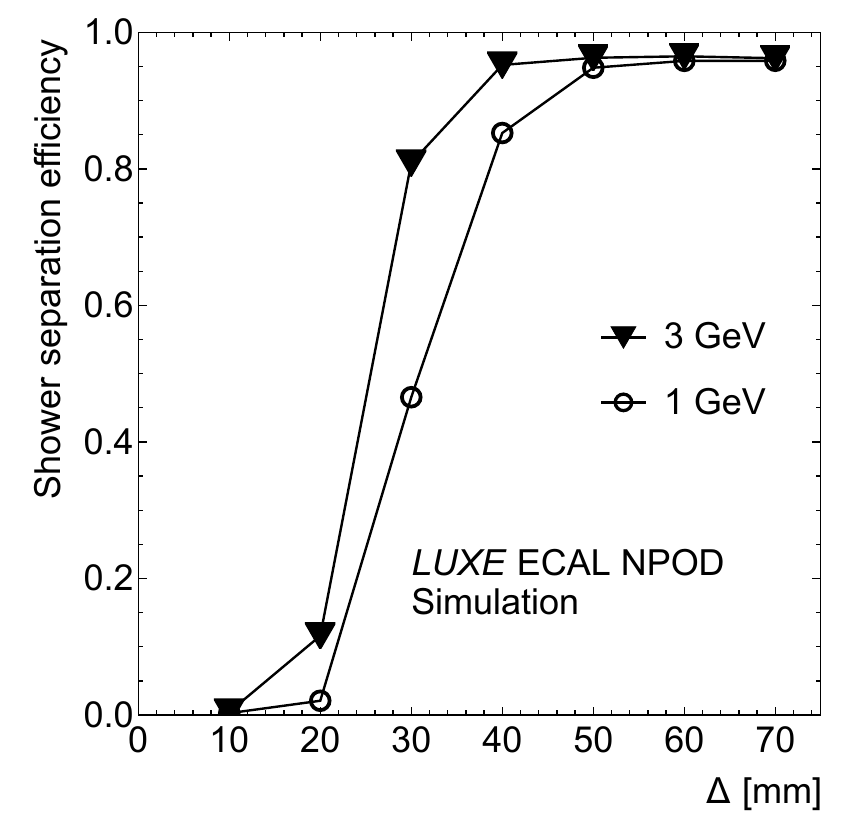}
    \caption{The separation efficiency by the nearest-neighbor clustering algorithm of photon pair showers against the distance $\Delta$ between the pairs at the entry point on the front surface of the detector.
    The simulations are done with 1 and 3~GeV photons.
    The solid interpolated lines between markers are used to suggest the trend across data points.
    }
    \label{fig:vertex_separation_power}
\end{figure}

As a proof-of-concept study, a well-established clustering algorithm called the nearest-neighbor (NN) clustering~\cite{Gaede_2007, Sikler_2010} is utilized.
This method connects hits that are within a specified distance from each other.
Other requirements, such as an energy deposit threshold, are also applied to reject isolated hits that are created by stray secondary photons.
The NN distance, deposit threshold and other clustering parameters are optimized by minimizing the difference between the reconstructed result and the simulation truth.
As the simulation uses two photons of the same energy, the clustering is considered successful
when the result gives exactly two major clusters and the ratio between their energy deposits is between 0.5 and 2.
Fig.~\ref{fig:vertex_separation_power} shows that when the two photons have a distance of 40 to 50~mm at the entry point on the front plane of the detector, the clustering algorithm can separate over 95\% of the simulated photon pairs for the energy range between 1 to 3~GeV.
The efficiency at smaller distances falls as expected and, e.g., goes down to 10\% at $20\cm$.
As shown in Fig.~\ref{subfig:phase1_sensitivity_minseparation}, setting a minimum di-photon separation of about 50 mm is sufficient for the current compact LUXE ECAL NPOD design.
When the distance is below $40\mm$,
the power of separation can be enhanced by incorporating more evolved clustering algorithms that exploit other topological information available in the compact calorimeter. For example, the characteristic shapes of electromagnetic showers can be used to disentangle the overlap. An implementation was shown in GARLIC~\cite{Jeans:2012jj}, and preliminary studies of applying template fitting in clustering have already shown improvements compared to the NNC. In addition,  approaches using particle-flow algorithms (PFA) could make use of the longitudinal development of the showers, particularly in the early layers before the overlap becomes significant. In the simulation of CEPC ECAL, Arbor~ \cite{Ruan_2014, Zhao:2017qcy} managed to separate two 5 GeV photons at the distance of 11 mm, around 2 times of their cell size. In recent years, the PFA methods are further enhanced by machine learning models, such as graph neural network, to improve their efficiency~\cite{Shpak_2018, Murata:2024xxe, Wahlen:2024rxt}.
The adaptation of them into our simulation workflow presents a perspective of future development.
These algorithms can also benefit from the adaptation of the hardware. Modifications to the calorimeter design, such as making the calorimeter more compact (by reducing the thickness of air gaps) or employing thinner absorber layers in the initial stages \cite{Borysov:2024kpa}, or altering the granularity of certain sensitive layers \cite{ALICE:2024jtt}, would improve the shower shapes and facilitate their separation.

\subsection{Photon kinematic reconstruction}
\label{subsec:recon_vertex}

\begin{figure}[!ht]
    \centering
    \begin{subfigure}[b]{0.495\textwidth}
        \centering
        \includegraphics[width=\textwidth]{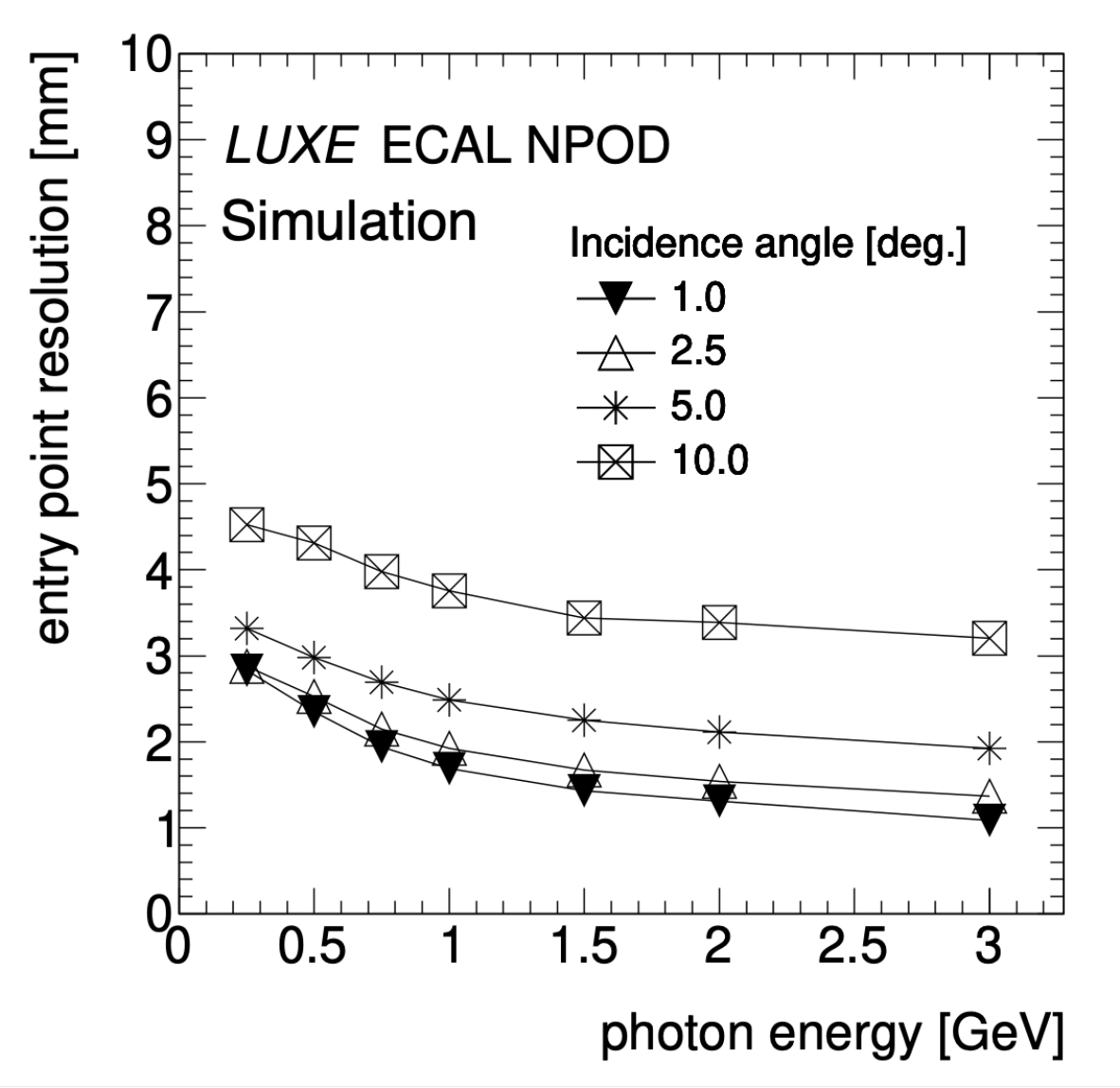}
        \caption{}
        \label{subfig:entry_point_res}
    \end{subfigure}
    \hfill
    \begin{subfigure}[b]{0.495\textwidth}
        \centering
        \includegraphics[width=\textwidth]{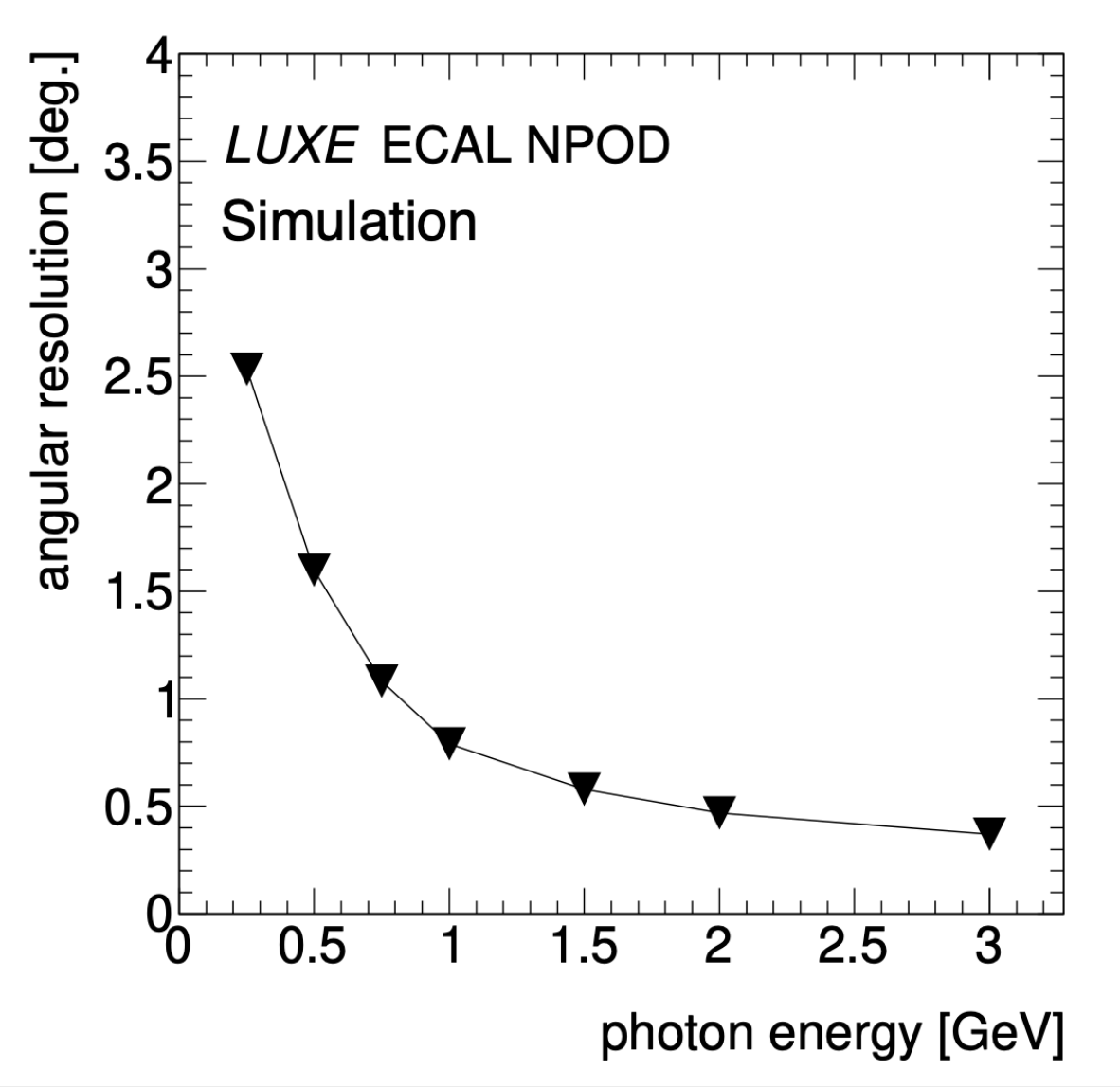}
        \caption{}
        \label{subfig:angle_res_onegraph}
    \end{subfigure}
    \caption{The estimated resolution for the entry point distance $\Delta$ on the front surface of the detector (a) and the angle of incidence (b) of single photons reconstructed in the ECAL NPOD.
    For the angular resolution estimation, a fixed $\Delta = 30 \mm$ is used, and there is no noticeable difference between the estimates with various incident angles.
    The solid interpolated lines between markers are used to suggest the trend across data points.
    }
    \label{fig:vertex_reconstruction_resolutions}
\end{figure}

Given that the showers are successfully separated by the clustering algorithm, 
each cluster is then reconstructed into a track through line-fitting among the hit points weighted by their energy deposits.
The total deposits of the cluster is used to reconstruct the photon energy,
and the track provides the photon's direction and its entry point at the calorimeter front surface.
The information is further used to estimate the ALP candidate's mass, lifetime, and the coupling constant.
To enhance the performance of the tracking, a re-clustering step is added, where the central cores of the NN clusters are used for the final tracking.
The relative size of the core is optimized to improve the tracking resolution.
The residuals of the reconstructed position at the front of the detector and the directional vector of a track are used to study the resolution of the tracking algorithm.
Different scenarios are compared in Fig.~\ref{fig:vertex_reconstruction_resolutions},
where it is shown that the current algorithm may provide resolution for the position of the entry point below 3~mm for most relevant energies and low incident angle, and below 5~mm for the largest incident angles.
For the incident angle, a resolution at the level of one degree and lower is achievable for most relevant energies with current algorithms.

\subsection{Particle identification and background rejection}
\label{subsec:recon_pid}
\begin{figure}[!ht]
    \centering
    \begin{subfigure}[b]{0.32\textwidth}
        \centering
        \includegraphics[width=\textwidth]{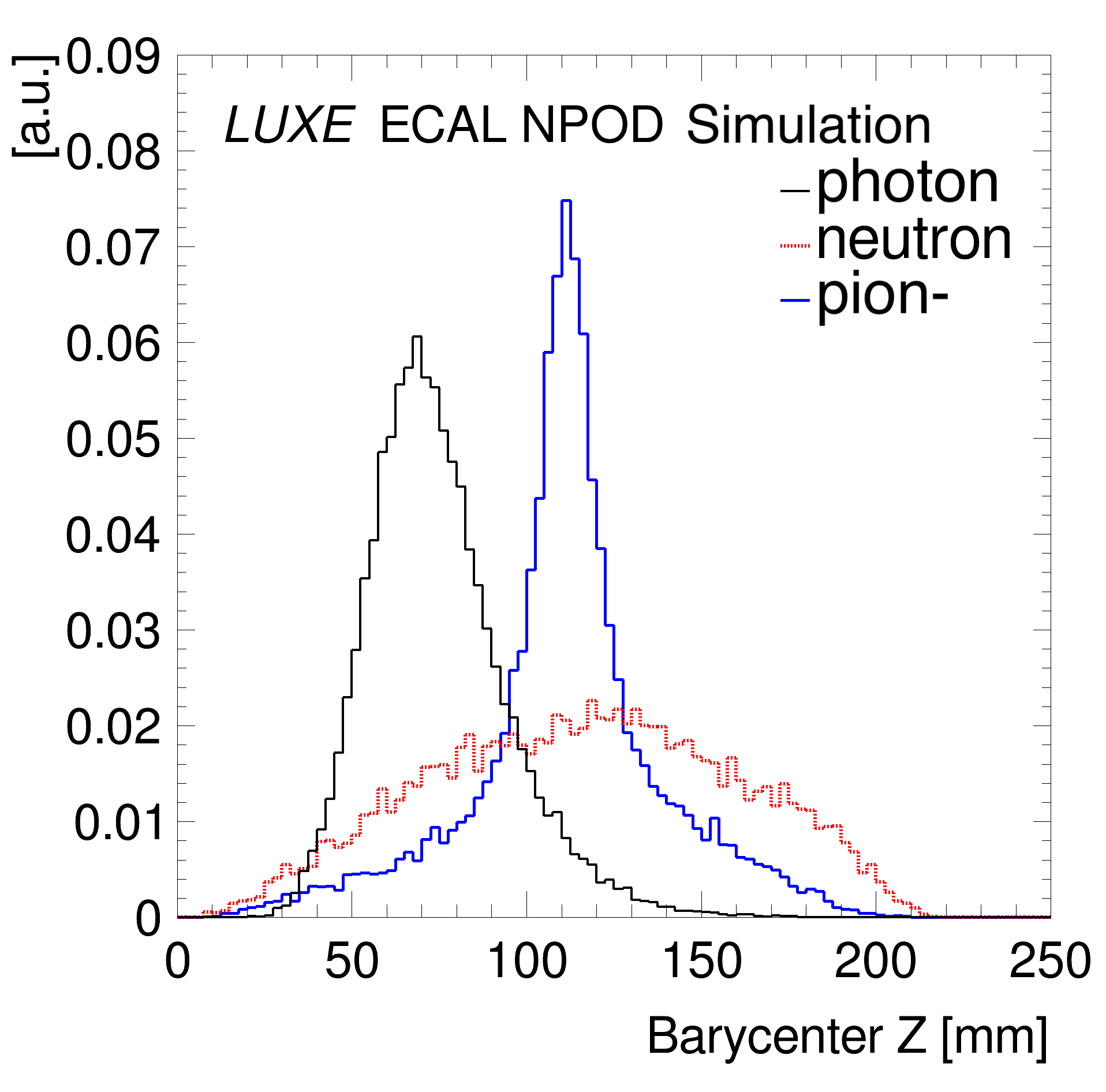}
        \caption{}
        \label{subfig:ecalnpod_pid_barz}
    \end{subfigure}
    \hfill
    \begin{subfigure}[b]{0.32\textwidth}
        \centering
        \includegraphics[width=\textwidth]{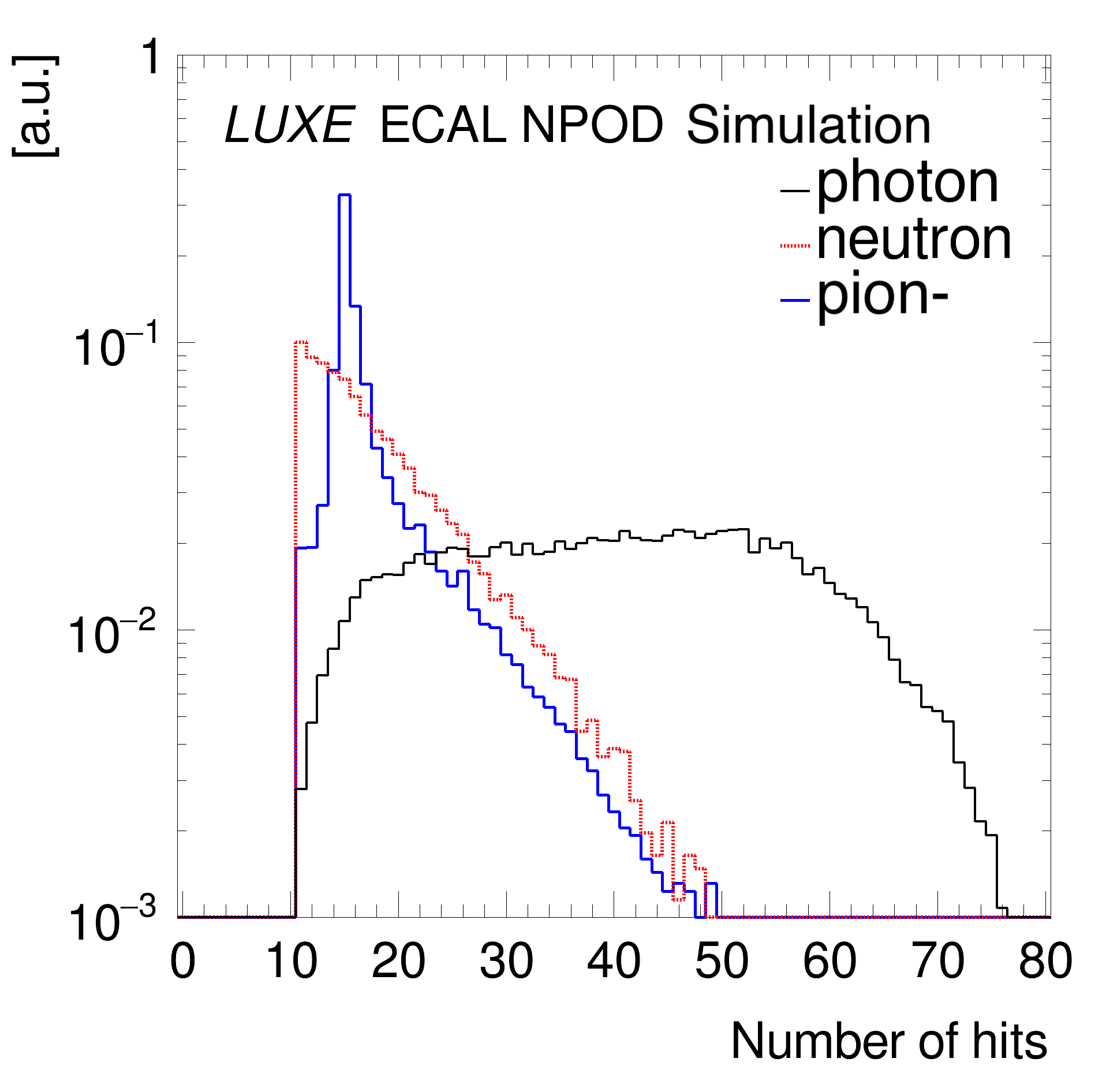}
        \caption{}
        \label{subfig:ecalnpod_pid_nhit}
    \end{subfigure}
    \hfill
    \begin{subfigure}[b]{0.32\textwidth}
        \centering
        \includegraphics[width=\textwidth]{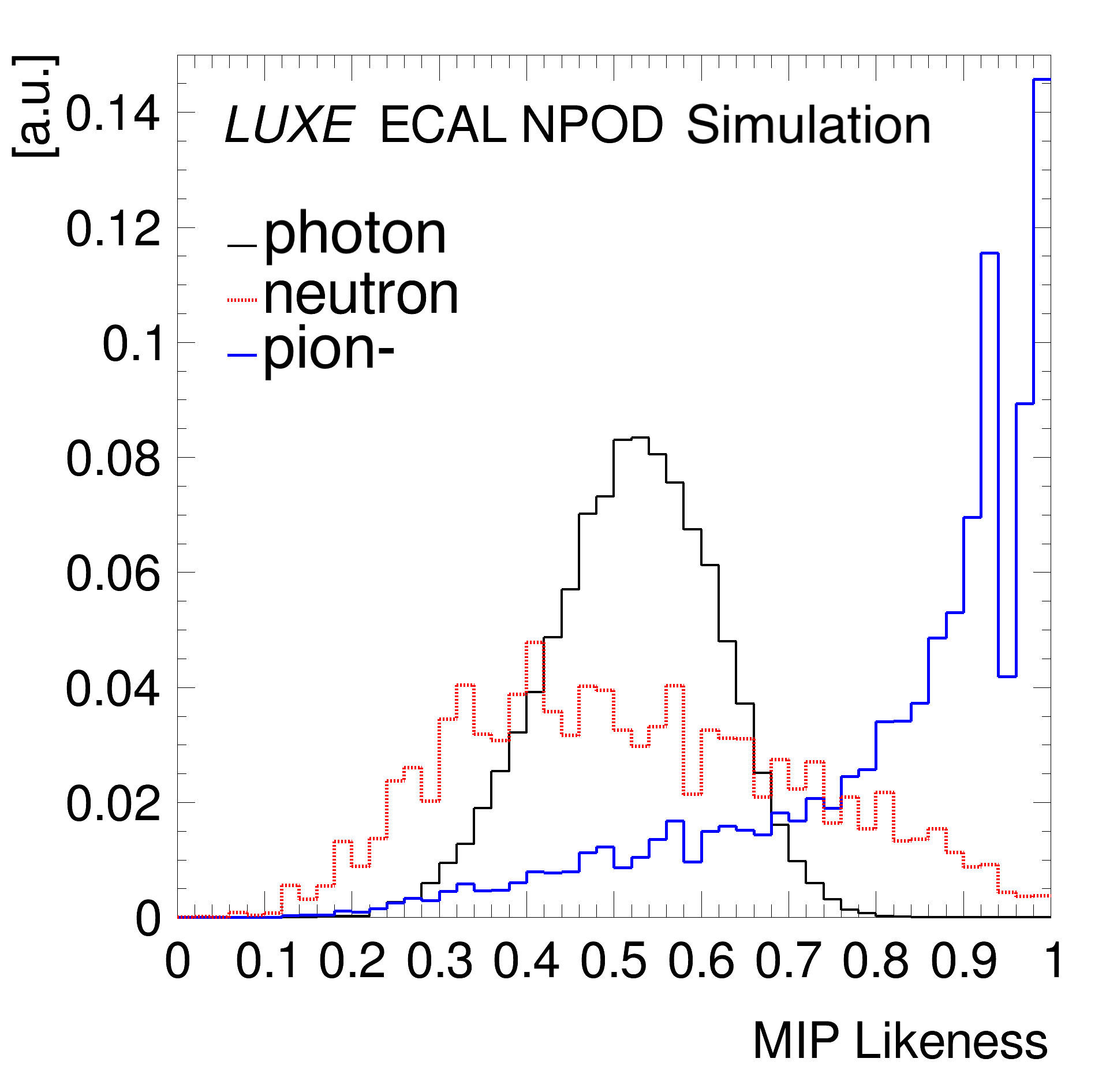}
        \caption{}
        \label{subfig:ecalnpod_pid_miplikeness}
    \end{subfigure}
    \caption{The distributions of three PID variables for the hits on the detector caused by incoming photon, neutron, and pion:
    the $z$ coordinate of the barycenter (a),
    the number (b), and
    the MIP likeness [defined in Eq. (\ref{eq:miplikeness})] (c) of the hits.
    Only the events where the particle deposits more than 10 hits in the detector are shown in the plots.}
    \label{fig:detector_pid_variables}
\end{figure}

The shower development inside a given material is different for photons and background particles, such as neutrons, pions, or protons.
For instance, in our simulations, over 46\% of the 0.5~GeV neutrons do not interact at all with the ECAL NPOD.
This fraction decreases slightly to below 44\% in the case of 2~GeV neutrons.
The remaining neutrons, by producing secondary particles, either generate a few MIP-like deposits or develop showers in random directions.
In contrast, most of the pions do not develop showers in the thin electromagnetic calorimeter.
Instead, they leave MIP-like hits along their trajectories.
The photon showers and the pion hits are aligned with the incident directions of the incoming particles.
These signatures, which are detected by a highly compact and granular calorimeter, can be parameterized and used to identify and reject background contamination.
Particle identification (PID) is used as one of the tools to reject background from pions, protons, or neutrons.
One of the PID methods is powered by a boosted decision tree (BDT) multivariate classifier~\cite{Roe_2005, Suehara_2016, CALICE_BDTPID_2020, JPMH_PhD_2025}. 
An implementation of this method in the Toolkit for Multivariate Data Analysis (TMVA)~\cite{Speckmayer_2010} in \textsc{Root}, which features a particle swarm optimization~\cite{Kennedy_1995} of hyper parameters, is used.
The detector responses are parametrized as several topological variables, including their barycenter,
transverse sizes, energy deposits, the resemblance to a minimum ionizing particle track, and the occupancy of the detector.
The resemblance of an event to a MIP track, called ``MIP likeness'', is calculated as
\begin{equation}
\label{eq:miplikeness}
    \text{MIP likeness}=\frac{1}{N_\text{Layer}}\sum^{N_\text{Layer}}_{n=1}{\frac{1}{S_n}},
\end{equation}
where $N_\text{Layer}$ is the total number of layers of the detector, and $S_n$ is equal to the number of hits in the $n$-th layer, if the layer has at least one hit, or $-1$ if the layer has no hits.
Fig.~\ref{fig:detector_pid_variables} shows examples of variable distributions for different incoming particles of photon, neutron, and pion.
A detailed list of all variables used in the BDT can be found at \cite{JPMH_PhD_2025}.
The training of the BDT associates the variable distribution patterns with the types of incoming particles.
It learns the importance of the variables as a measure to distinguish different particles.
After the first iteration, highly correlated variables are removed from the BDT training.

\begin{figure}[!ht]
    \centering
    \includegraphics[width=0.495\textwidth]{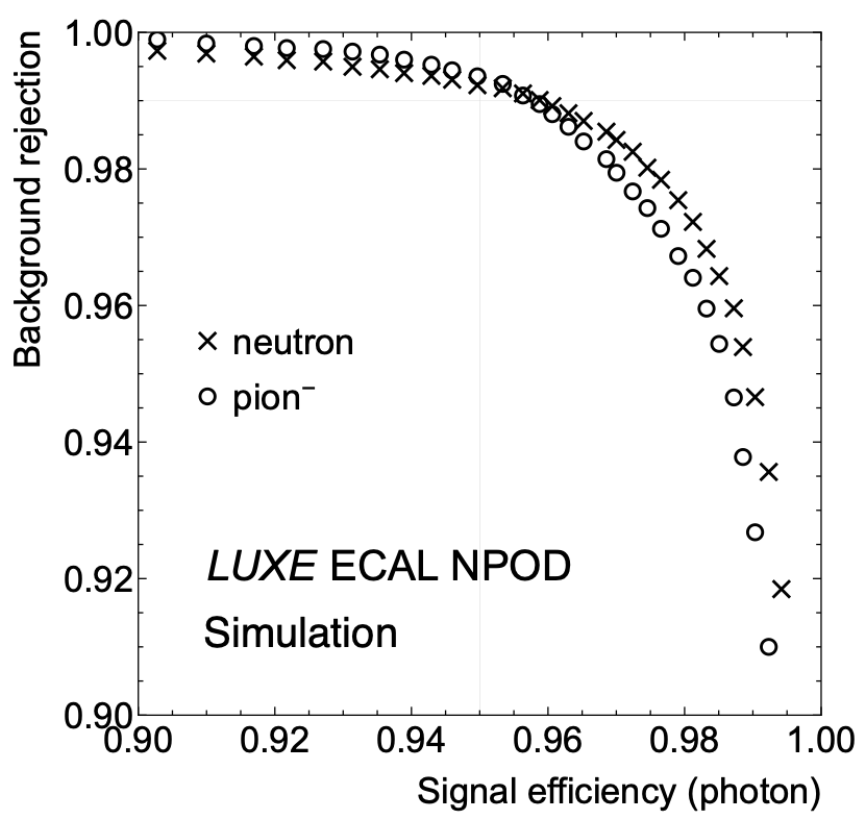}
    \caption{Background rejection rates against photon efficiency in a BDT model trained with photon, pion, and neutron.
    }
    \label{fig:pid_photon_pion_neutron}
\end{figure}

The simulated dataset for the BDT training contains $10^4$ particles per kind (photons, pions, and neutrons).
The particles have a flat kinetic energy spectrum between $0.5\GeV$ and $3 \GeV$, which is of interest for the NPOD case.
Each kind of particle is treated by the BDT model as an individual category in the training.
\begin{figure}
    \centering
    \includegraphics[width=0.495\linewidth]{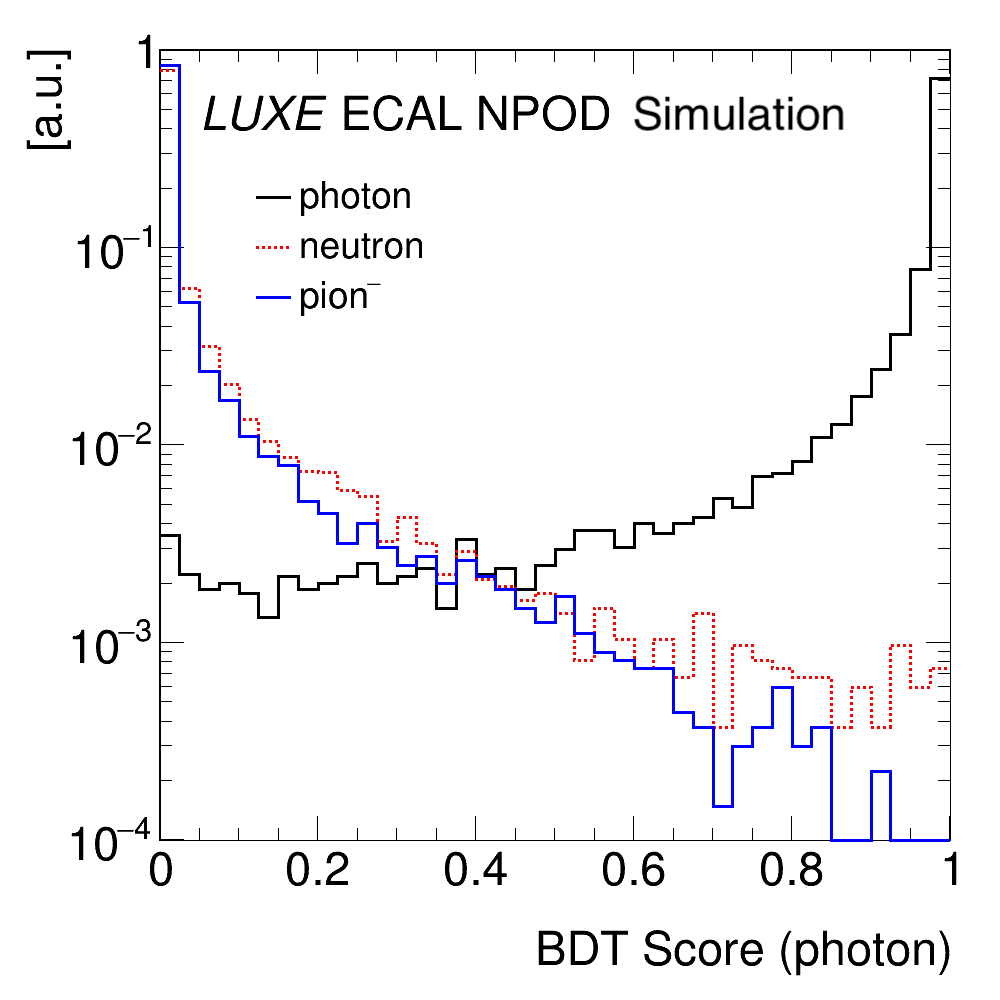}
    \caption{BDT score distributions for photon, neutron, and pion samples.
    }
    \label{fig:pid_photon_likelihood}
\end{figure}
After training, the model is fed with events of the detector's responses to a single particle.
Each event is given a score related to the probability of being signal. The BDT score distributions for the three categories are shown in Fig.~\ref{fig:pid_photon_likelihood}. 
Those events that have a score higher than a given threshold are considered as signal events, and the rest are rejected as background.
The fraction of reconstructed signal events in total simulated signal events is defined as signal efficiency
while the ratio between the rejected background events and the simulated background events is called background rejection efficiency.
A higher score threshold will reject more background events at the cost of losing some signal events.
And a lower score threshold can result in more signal events, but the background contamination will increase accordingly.
The performance of the BDT trained considering photons as signal and pions and neutrons as background is shown in Fig.~\ref{fig:pid_photon_pion_neutron}.
For a background rejection ratio at 99\%, this trained model provides a signal efficiency above 95\%.
For background from other charged particles, muons can be excluded based on their MIP-like signals in a calorimeter, while protons can be distinguished by the BDT in a similar way as for pions and protons described in this section.
More details are put in Appendix \ref{app:bdt_proton}.

\subsection{Summary of the detector impact}

With a transverse cross section of only $36\times18 \cm^2$, the ECAL-NPOD provides large enough acceptance for NP particle sensitivity (as discussed in Sec.~\ref{sec:npodsensitivity}).
The detector receives much lower background comparing to the estimate discussed in Sec.~\ref{sec:background} with a 1~m radius generic detector.
The same detector also provides the necessary timing resolution such that further residual backgrounds can be removed.
Lastly, $\sim 99$\% of the remaining background particles can be removed by the BDT, while preserving over $\sim 95$\% of signal.

The signal di-photon system can be fully resolved if the photons are separated by 40 to 50~mm, and even below that albeit with lower efficiency.
The tracking algorithm provides position resolutions of 1 to 3~mm for small photon incident angles below 5 degrees and 3 to 5~mm for larger angles.
The directional resolutions are better than 1~degree for photons above 1~GeV, and between 1 to 2~degrees around 0.5~GeV.
This performance enables to infer the vertex of the signal pair and to reconstruct the mass and coupling constant of the NP particle.

Taken together, we conclude that with the existing ECAL-E as the NPOD detector, this search can be regarded as being background free.
This agrees with the results of the previous study~\cite{Bai_2022}, but now relies on a much more realistic experimental basis.

\section{Results}
\label{sec:results}
Our dump design considers two fundamental aspects: the NP production and background suppression.
For phase-0.10/40, where the background is expected to be significantly smaller than for phase-1, a shorter tungsten core dump of length 25~cm and radius 10~cm is sufficient to reach a background-free environment. 

On the other hand, for phase-1, as the background is non-negligible before considering the detector, a comprehensive study is performed for different dump materials and sizes.
It is found that neutrons are generated in particular from scattering with elements of the experiment setup, leading to the need of a dump with large transverse area. 
As a result, the most adequate dump for phase-1 is a core tungsten dump with length 1~m and radius 20~cm, wrapped in an outer dump made of lead of length 1~m and outer radius 50~cm.

The decay volume length is also optimized with particular attention to the parameter space that LUXE-NPOD aims at probing and the detector design.

Finally, combined with the dump design for phase-1, we show that the silicon-tungsten ECAL LUXE NPOD sampling calorimeter, can provide the necessary background rejection power for a background-free search, while also preserving a high enough signal efficiency for viable signal parameter space.
The detector performance is also promising in terms of the ALP reconstruction.

\begin{figure}[!ht]
\centering\includegraphics[width=0.745\textwidth]{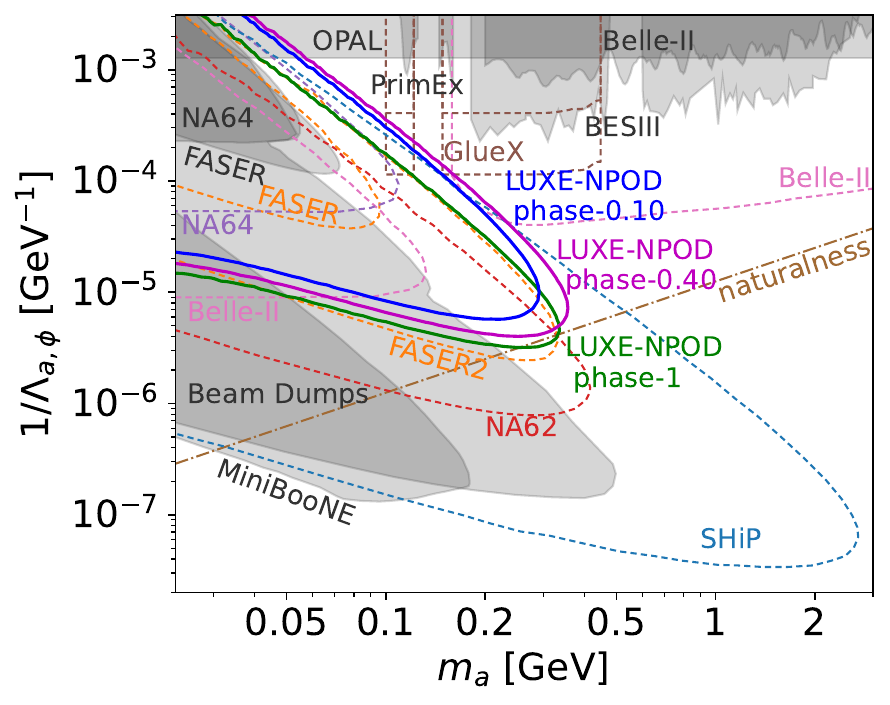}
\caption{The projected reach of LUXE-NPOD phase-0.10 at $10\TW$, phase-0.40 at $40\TW$, and phase-1 at $350\TW$. Currently existing bounds (gray regions) NA64, PrimEx, OPAL, Belle~II, BESIII, FASER, MiniBoone and beam-dumps~\cite{PhysRevLett.125.081801, PhysRevLett.123.071801, PhysRevLett.118.171801, PhysRevLett.125.161806, Ablikim_2023, PhysRevD.110.L031101, BERGSMA1985458, PhysRevLett.59.755, Dolan_2017, na62_dump, PhysRevD.108.075019, faser_2025} and future projections (dashed lines) from Belle~II, NA62-dump, GlueX, FASER, FASER2, NA64 and SHiP~\cite{Dolan_2017, na62_dump, PYBUS2024138790, PhysRevD.102.055018,Albanese:2878604}.}
\label{fig:npod_sensitivity}
\end{figure}

Assuming zero background, the projections for the sensitivity of LUXE-NPOD phase-0.10/40
(tungsten dump, $L_D=0.25 \m$, $L_V=2.5 \m$, $R_\text{Det}=1 \m$, no minimum photon separation requirement) and phase-1 (tungsten dump, $L_D=1.0 \m$, $L_V=1 \m$, $R_\text{Det}=1 \m$, no minimum photon separation requirement) 
are presented in Fig.~\ref{fig:npod_sensitivity}. 

We compare the results with the current bounds (gray regions) of NA64~\cite{PhysRevLett.125.081801}, PrimEx~\cite{PhysRevLett.123.071801}, OPAL~\cite{ PhysRevLett.118.171801}, Belle-II~\cite{PhysRevLett.125.161806}, BESIII~\cite{Ablikim_2023, PhysRevD.110.L031101}, FASER~\cite{faser_2025}, MiniBoone~\cite{PhysRevD.108.075019} and beam-dumps~\cite{BERGSMA1985458, PhysRevLett.59.755, Dolan_2017, na62_dump}. The dashed lines are future projections of Belle~II~\cite{Dolan_2017}, NA62-dump~\cite{na62_dump}, GlueX~\cite{PYBUS2024138790},  FASER~\cite{PhysRevD.98.055021}, 
FASER2~\cite{PhysRevD.98.055021}, NA64~\cite{PhysRevD.102.055018} and SHiP~\cite{Albanese:2878604}.

Thanks to the short dump design, LUXE-NPOD phase-0.10 can already probe the unexplored region of the parameter space in the mass range of $35\MeV \lesssim \mx \lesssim 300\MeV$, and 
$1/\Lambda ~>~ 1\times 10^{-5}$.
LUXE-NPOD phase-0.40 may probe the mass range of $35 \MeV \lesssim \mx \lesssim 350 \MeV$, and 
$1/\Lambda ~>~ 6\times 10^{-5}$.
With a longer dump, LUXE-NPOD phase-1 can probe the region of the parameter space in the mass range of $27 \MeV \lesssim \mx <~ 330 \MeV$, and $1/\Lambda ~>~ 4\times 10^{-6}$.
The natural parameter space for $CP$ even scalar is below the brown dashed-dotted line, and is defined in \cite{Bai_2022}.

\section{Outlook}
\label{sec:outlook}

In this work we perform a detailed study of the LUXE-NPOD proposal from \cite{Bai_2022}. We adopt more realistic assumptions, more simulated data, and a concrete strategy for the di-photon reconstruction and background rejection.
With a projection to become online in 2030, this work concretely positions LUXE-NPOD as a unique and major player in the landscape of beam dump experiments and ALPs search.

The study presented is carried out considering the current conditions of the LUXE experimental setup, leaving room for further optimizations.
Particularly, the background coming from charged particles has been analyzed, but the study of a design including a magnet between the dump and the detector to suppress this background (beyond what can be achieved with the PID algorithms) pends on the decision of the experiment location.
In addition, the preliminary study of calorimeter performance can be improved with dedicated reconstruction algorithms.

\section*{Acknowledgements}

The authors would like to thank Tom Blackburn for help with \PTARMIGAN. 
We are also grateful to the SiW-ECAL team of CALICE and the DRD6 collaborations for the useful discussions. 
KIT members acknowledge the support of the Alexander von Humboldt Foundation.
IFIC members acknowledge the financial support
from: the Spanish MICIU/AEI and European Union/FEDER via the grant \texttt{PID2021-122134NB-C21} the Generalitat Valenciana (GV) via the Excellence Grant \texttt{CIPROM/2021/073} and the PlanGenT program with the grant number \texttt{CIDEGENT/2020/021}; the MCIN with funding from the European Union NextGenerationEU and Generalitat Valenciana in the call ``Programa de Planes Complementarios de I+D+i (PRTR 2022)'' through the project with reference \texttt{ASFAE/2022/015}; and the Program Programa Estatal
para Desarrollar, Atraer y Retener Talento PEICTI 2021-2023 through the project
with reference \texttt{CNS2022-135420}.
IFIC members also acknowledge the computer resources at Gluon and Artemisa.
Artemisa is funded by the European Union ERDF and Comunitat Valenciana as well as the technical support provided by IFIC, CSIC-UV.
The work of Noam Tal Hod's group at Weizmann is supported by a research grant from the Estate of Dr.\ Moshe Gl\"{u}ck, the Minerva foundation with funding from the Federal German Ministry for Education and Research, the Israel Science Foundation (grant No.\ 708/20 and 1235/24), the Anna and Maurice Boukstein Career Development Chair, the Benoziyo Endowment Fund for the Advancement of Science, the Estate of Emile Mimran, the Estate of Betty Weneser, a research grant from the Estate of Gerald Alexander, a research grant from the Potter's Wheel Foundation, a research grant from Adam Glickman and the Sassoon \& Marjorie Peress Legacy Fund, the Deloro Center for Space and Optics, and in part by the Krenter--Perinot center for High-Energy particle physics.
Yotam Soreq is supported by grants from the NSF-BSF (grant No.\ 2021800) and from ISF (grant No.\ 597/24). 

\printbibliography


\clearpage
\appendix
\section{Magnetized dump studies}
\label{app:magnetized_dump}

\begin{figure}[!ht]
    \centering
    \includegraphics[width=0.8\textwidth]{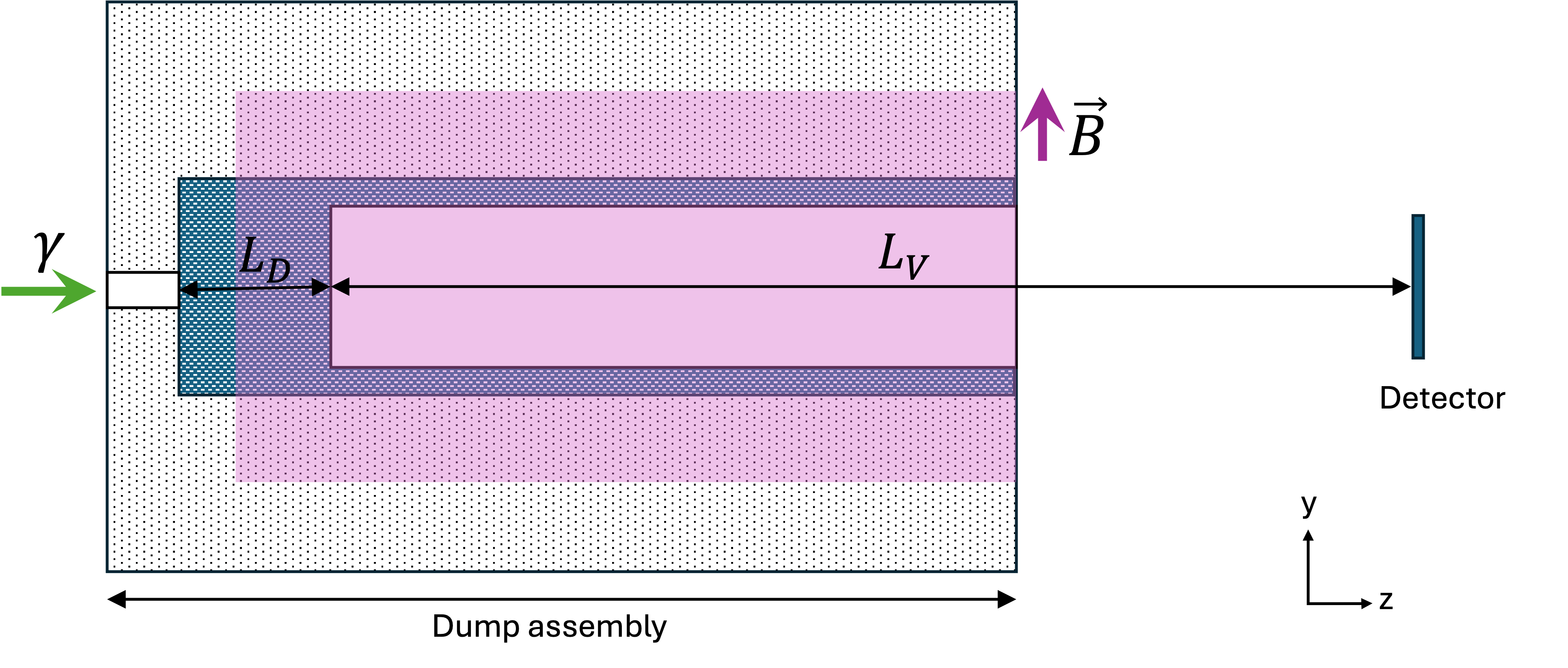}
    \caption{A 2D diagram of magnetized dump design. The solid (dark blue) cylinder is the tungsten dump core with length $L_D$.
    The outside cylinder (gray) is the lead dump wrap.
    The pink shadowed area is the magnetized iron wrap.
    The concrete wall is not shown.
    The slit at the face of the dump helps to reduce the backscattered component and it is present in all designs used in this work.
    The diagram is not to scale.}
    \label{fig:magnetized_dump_diagram}
\end{figure}

Aiming for a shorter dump design, we also explore the possibility of implementing a magnetized dump.
This kind of solution was proposed, e.g., in~\cite{VANELP1997403,March:1971ps}.
A magnetic field can deflect the charged particles (electrons, protons, pions, etc.) in the electromagnetic and hadronic showers produced by the incoming photon beam.
In this way the neutrons will be produced with a larger angle compared to the forward direction and, naively speaking, will not reach the detector.
To avoid interfering with the NP production occurring within the first mm's of the dump, we propose a magnetized iron wrap around the tungsten core, which begins a few mm's after the beginning of the dump.
This setup is shown in Fig.~\ref{fig:magnetized_dump_diagram}.
We do not comment on the technical feasibility of this proposal, but only focus on the question whether it can be beneficial for defocusing the showers and hence potentially reducing the collinear backgrounds.

\begin{table}[!ht]
\centering
\small\setlength\tabcolsep{4.5pt}
\begin{tabular}{cccccccc}
\hline
 Setup, $N_{\rm BX}$ &
  $B$ [T] &
  $L_D$ [cm] &
  $L_V$ [cm] &
  \begin{tabular}[c]{@{}c@{}}$N_\gamma$ \\ all \end{tabular} &
  \begin{tabular}[c]{@{}c@{}}$N_\gamma$ \\ $E_\text{kin} > 0.5 \GeV$\end{tabular} &
  \begin{tabular}[c]{@{}c@{}}$N_n$\\ all \end{tabular} &
  \begin{tabular}[c]{@{}c@{}}$N_n$ \\ $E_\text{kin} > 0.5 \GeV$\end{tabular} \\
\hline
a, 0.2 & 0   & 20 & 180 & 2900 & 0 & 170000 & 7500 \\
b, 0.2 & 1.5 & 20 & 180 & 2500 & 0 & 170000 & 7500 \\
c, 0.2 & 1.5 & 50 & 150 & 0 & 0 & 23000 & 680 \\
d, 0.2 & 5.0 & 50 & 150 & 530 & 0 & 24000 & 380 \\
\hline
\end{tabular}
\caption{Number of background photons ($N_\gamma$) and neutrons ($N_n$) arriving at the detector per BX.
The dump is enclosed in an iron magnetized wrap with a decay volume $L_V=1\m$. The detector used for the simulations is LUXE ECAL-E (Sec.~\ref{subsec:detector_candidate}).
Different dump lengths and magnetic fields are tested.
A total of 0.2 BXs are simulated for each setup, and the numbers are normalized to 1 BX.}
\label{table:phase1optimizationmagneticbkg}
\end{table}

Table~\ref{table:phase1optimizationmagneticbkg} shows the results of different magnetized dump parameters. As the main goal of proposing a magnetized dump setup is to keep the experiment background-free, while using a short dump, we first simulate $L_D=0.2\m$ with no magnetic field (Table~\ref{table:phase1optimizationmagneticbkg}.a) and a magnetic field of $1.5 \T$ (Table~\ref{table:phase1optimizationmagneticbkg}.b) respectively.
No significant gain in background suppression is seen by adding a magnetic field.
Moreover, there are $\mathcal{O}(10^3)$ neutrons per BX above $E_\text{kin}=0.5 \GeV$ in both cases.
A longer dump $L_D=0.5 \m$ with magnetic field of $1.5 \T$ (Table~\ref{table:phase1optimizationmagneticbkg}.c) 
is therefore tested. We see that there are still $\mathcal{O}(10^2)$ neutrons remaining per BX above $E_\text{kin}=0.5 \GeV$.
Finally, the magnetic field is increased to $5 \T$ for a dump of $L_D=0.5 \m$ (Table~\ref{table:phase1optimizationmagneticbkg}.d) and still there are $\mathcal{O}(10^2)$ neutrons seen per BX above $E_\text{kin}=0.5 \GeV$.
Since there is no gain seen on background reduction, and given the technical complexity associated with this setup, we chose the W+Pb dump configuration (Table~1.g) as the most efficient for LUXE phase-1.

\section{Background rejection for other charged particles}
\label{app:bdt_proton}
Section \ref{sec:background} sees the background from neutrons, pions, protons, muons, and electrons.
Assuming a deflective magnetic field can get rid of the charged particles, in Sec. \ref{subsec:recon_pid}, only the most significant background from neutrons and pions has been discussed.

\begin{figure}[!ht]
    \centering
    \includegraphics[width=0.495\textwidth]{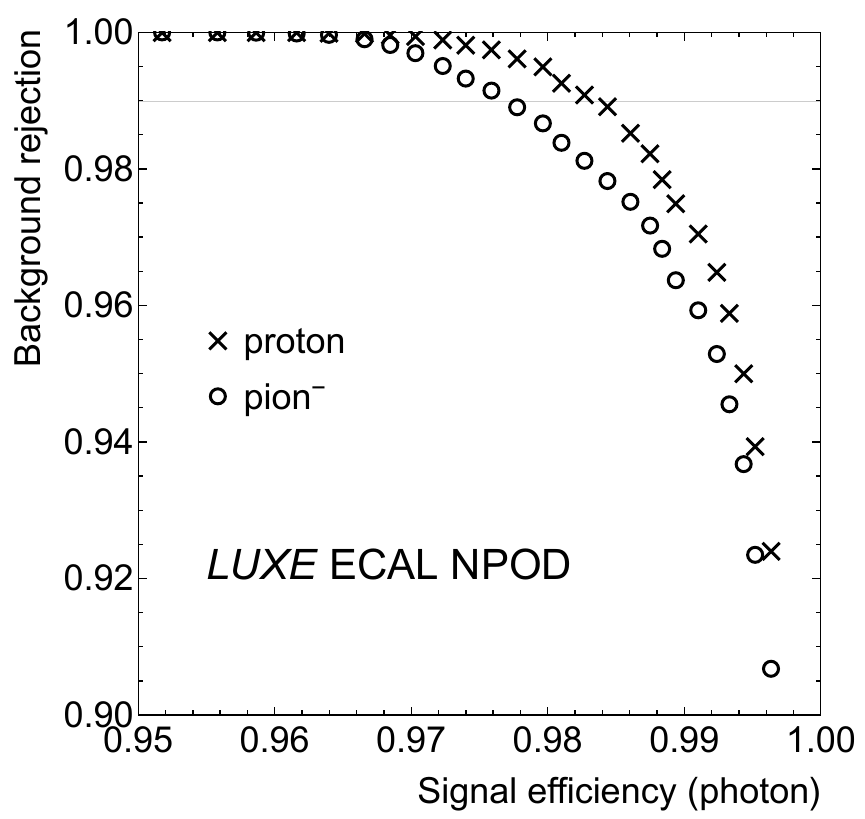}
    \caption{Background rejection rates against photon efficiency in a BDT model trained with photon, pion, and proton.
    Compared to Fig.~\ref{fig:pid_photon_pion_neutron}, the $x$ axis here starts from 0.95 rather than 0.90.
    Different particle combinations in the training datasets account for the difference in pion rejection rates.
    }
    \label{fig:pid_photon_pion_proton}
\end{figure}

Though the presence of a magnet would offer additional background suppression margin, in the case of the absence of a sweeping magnet, we would like to point out that the design of LUXE NPOD has the ability to handle the residue of charged particle background, which is supported by the following:
(1) Muons only leave MIP-like signals in the calorimeter. Before use of the BDT, they can be readily excluded.
(2) Electrons are more challenging to separate from photons in an electromagnetic calorimeter alone. However, as shown in Fig. \ref{subfig:charged_energy}, their energies are below 0.5 GeV, and are not considered as potential signals.
(3) For protons, we have trained a BDT using photon, pion, and proton as inputs, and it achieves a  better background rejection rate (Fig. \ref{fig:pid_photon_pion_proton}) at the same signal efficiency compared to the photon–pion–neutron configuration.
Nonetheless, we cannot state at this stage if it can make the magnet completely redundant.

\end{document}